\documentclass{aa} 

\usepackage{graphicx}
\usepackage{supertabular}
\usepackage{txfonts, mathrsfs}
\usepackage{color}
\usepackage{times}
\usepackage[draft]{hyperref}
\usepackage{amsmath}
\usepackage{amssymb}
\usepackage{xcolor}
\usepackage{url}
\usepackage{natbib}

\begin{document}
\title{Tight asteroseismic constraints on core overshooting and diffusive mixing in 
the slowly rotating pulsating B8.3V star KIC\,10526294}

\author{E.~Moravveji\inst{1,}\thanks{Postdoctoral Fellow at KU\,Leuven 
funded by the Belgian Federal Science Policy Office (BELSPO).}
\and
C.~Aerts\inst{1,2}
\and
P.~I.~P\'{a}pics\inst{1,}\thanks{Postdoctoral Fellow of the Fund for Scientific 
Research (FWO), Flanders, Belgium.}
\and
S.~A.~Triana\inst{1}
\and
B.~Vandoren\inst{1}
}

\institute{Institute of Astronomy, KU\,Leuven, Celestijnenlaan 200D, B-3001
  Leuven, 
Belgium\\
\email{Ehsan.Moravveji@ster.kuleuven.be}
\and
Department of Astrophysics, IMAPP, Radboud University Nijmegen, 
PO Box 9010, 6500 GL, Nijmegen, The Netherlands\\
}

\titlerunning{Core overshooting and diffusive mixing in KIC\,10526294}
\authorrunning{E.\ Moravveji et al.}

\date{Received ...; accepted ...}

\abstract {KIC\,10526294 was recently discovered to be a very slowly rotating and
  slowly pulsating late B-type star.  Its 19 consecutive dipole gravity modes
  constitute a series with almost constant period spacing.  This unique
  collection of identified modes probes the near-core environment of this star
  and holds the potential to reveal the size and structure of the overshooting
  zone above the convective core, as well as the mixing properties of the  star.}
{We revisit the asteroseismic modelling of this star with specific emphasis on
  the properties of the core overshooting, while considering additional
  diffusive mixing throughout the radiative envelope of the star.}
{We pursued forward seismic modelling based on adiabatic eigenfrequencies of
  equilibrium models for eight extensive evolutionary grids tuned to
  KIC\,10526294 by varying the initial mass, metallicity, chemical mixture, and
  the extent of the overshooting layer on top of the convective core.  We
  examined models for both OP and OPAL opacities and tested the occurrence of extra
  diffusive mixing throughout the radiative interior.}
{ We find a tight mass-metallicity relation within the 
ranges $M\in [3.13,3.25]$\,M$_\odot$ and $Z\in [0.014,0.028]$. We deduce
that an exponentially decaying diffusive core overshooting prescription
  describes the seismic data better than a step function formulation and derive a value of
  $f_{\rm ov}$ between 0.017 and 0.018. Moreover, the inclusion of extra
  diffusive mixing 
  with a value of $\log D_{\rm mix}$ between 1.75 and 2.00\,dex 
  (with $D_{\rm mix}$ in \,cm$^2$\,s$^{-1}$) improves the goodness-of-fit based on 
  the observed and modelled frequencies by a factor $\sim\!11$ compared to the case 
  where no extra mixing is considered, irrespective of the $(M,Z)$ combination
  within the allowed seismic range.}
{We find that the inclusion of diffusive mixing in addition to core overshooting
  is essential to explain the structure in the observed period spacing pattern of this
  star.  Moreover, for the input physics and chemical mixtures we investigated,
  we deduce that an exponentially decaying prescription for the core
  overshooting is to be preferred over a step function, 
  regardless of the adopted mixture or choice of opacity tables.
  Our best models for KIC\,10526294 approach the seismic data to a level that
  they can serve future inversion of its stellar structure.
  	}


\maketitle

\section{Introduction}
\label{s-intro}

Intermediate and massive ($M\gtrsim2.5$ M$_\odot$) main-sequence (MS) stars of
spectral type OB harbour a fully developed convective core below a radiative
envelope during their core hydrogen-burning phase.  As their MS evolution
progresses, the decrease in the central hydrogen content $X_{\rm c}$ implies
that the convective core shrinks and it hence leaves behind a chemical gradient
$\nabla_\mu$ of CNO-processed material at the interface between the inner
convective core and the outer radiative envelope.  The extent of this interface
zone and its mixing properties remain largely unknown.  Asteroseismology in
principle offers the opportunity to probe this zone.

Stars are multi-component fluids, consisting of chemical species $i$ with a mass 
fraction denoted as $X_i$.
The spatial abundance distribution $X_i(r,t)$ does not stay constant during 
the evolution.
In burning regions, thermonuclear reactions provide a balance between the depletion 
of the elements $i$ that are burnt and the production of heavier elements, 
while the chemical mixture stays intact in the envelope.
In the absence of rotation and in a diffusive approximation, 
the spatial and temporal change 
of the individual mass fractions $X_i$ are governed by the following transport
equation in the Lagrangian description \citep[e.g.][]{heger-2000-01}:
\begin{equation}\label{e-Xi}
\frac{dX_i}{dt} = \left(\frac{dX_i}{dt}\right)_{\rm nuc} +
                  \frac{\partial }{\partial m}\left[(4\pi r^2\rho)^2 \, D_{\rm mix} \, 
                  \left(\frac{\partial X_i}{\partial m}\right) \right],
\end{equation}
where the first term accounts for the changes induced by nuclear burning, while 
the second 
term accounts for the mixing of element $i$ according to its 
spatial gradient $\partial X_i/\partial m$. In this equation,
$D_{\rm mix}$ is the effective diffusive mixing coefficient, which
has a unit of 
area over time \citep[e.g.][]{maeder-2009-book}.
Advective phenomena such as meridional circulation are excluded in this formulation.
Equation\,(\ref{e-Xi}) is solved subject to boundary conditions for the abundance
gradients in the core and at the surface, that is, 
$(\partial X_i/\partial m)_{\rm core}=(\partial X_i/\partial m)_{\rm surface}=0$.

A collection of mechanisms contributes to the local mixing in the stellar interior.
Convection is the dominant one, because it homogenises convective zones on a
dynamical timescale, which is very short compared to the nuclear timescale.
Mixing in the radiative interior occurs on a much longer timescale.
A handful of mixing mechanisms that can operate in the stellar envelope have
been proposed, such as core convective overshooting
\citep{saslaw-1965-01, roxburgh-1965-01, shaviv-1973-01}, semi-convective mixing
\citep{schwarzschild-1958-01, kato-1966-01, langer-1983-01}, rotational mixing
\citep{weiss-1988-01,charbonnel-1992-01,chaboyer-1994-01}, shear-induced mixing
\citep{jeans-1928-book, zahn-1992-01}, and mixing induced by internal
magnetic fields \citep{heger-2005-01}. Following a multivariate analysis of 
the seismic, magnetic, and rotational properties of a
sample of slowly rotating 
Galactic OB-type stars, \citet{aerts-2014-01} found the observational
evidence for  pulsationally induced mixing to be stronger than for rotational mixing.

In the computation of stellar evolution models, most of these mixing mechanisms
are effectively defined in a diffusion approximation and described by a
diffusion coefficient \citep[e.g.][]{heger-2000-01}.
In the absence of quantitative observational constraints on the individual
contributions for each of the mixing mechanisms, their net effect $D_{\rm mix}$ is
either linearly summed up \citep{heger-2000-01, heger-2005-01} or their
interaction is considered \citep{maeder-2013-01, ding-2014-01}.
In this paper, $D_{\rm mix}$ in radiative interior is a linear superposition of individual mixing processes.
Any local modification to the shape of the mixing profile $D_{\rm mix}(r)$ propagates
into the stellar structure through Eq.\,(\ref{e-Xi}) and subsequently influences the
shape of the Brunt-V{\"a}is{\"a}l{\"a} frequency, which incorporates the gradient
of the mean molecular weight.
Given the sensitivity of high-order g-modes to the detailed shape of the 
Brunt-V{\"a}is{\"a}l{\"a} frequency, their frequencies are influenced by
the local shape of $D_{\rm mix}(r)$.
For this reason, and thanks to asteroseismology based on the high-precision 
{\it Kepler\/} space-based 
photometry, we are able to investigate whether 
observational constraints on $D_{\rm mix}$ are within reach.

Before high-precision space photometry was available,
\citet{miglio-2008-01} have already shown how varying $D_{\rm mix}$
influences the morphology of the period spacings of gravity modes in
MS stars with convective cores.
\citet{degroote-2010-01} applied this approach to model the detected period
  spacing of the slowly rotating pulsating B3V star HD\,50230 observed with the
  CoRoT mission for five months.
  They assumed the detected sequence of consecutive modes belongs to the 
  dipole ($\ell=1$) series.  
  They concluded that HD\,50230 has a mass
  between 7 and 8\,M$_\odot$, has consumed some 60\% of its initial hydrogen,
  and required a value of $\log D_{\rm mix}$ between 3.48 and
  4.30\,cm$^2$\,s$^{-1}$ to explain the small periodic deviation of 240\,s from
  a constant period spacing of 9\,418\,s for an assumed value of 0.02 for the
  metallicity.  However, this result was criticised by \citet{szewczuk-2014-01}, who
  suggested that some of the low-amplitude peaks in the series might be due to
  modes of different degree.

KIC\,10526294 was recently discovered to be a very slowly rotating and slowly
pulsating B (SPB) star from {\it Kepler} photometry \citep[][hereafter
P14]{papics-2014-01}, exhibiting 19 rotationally split dipole gravity modes.
P\'{a}pics and collaborators
 performed forward seismic modelling and considered core
overshooting in the diffusive exponentially decaying prescription defined in
\citet{freytag-1996-01} and \citet{herwig-2000-01}.  We designate the free
parameter of this prescription as $f_{\rm ov}$ (and discuss it in
Eq.\,(\ref{e-exp-ov}) below).  But, they only succeeded to put an upper limit on
this free parameter: $f_{\rm ov}\lesssim\,0.015$.  

In contrast to the case of HD\,50230, there is no doubt about the
  identification of the degree of the modes for KIC\,10526294.
Hence, we here revisit the forward modelling of this SPB
(Sect.\,\ref{s-kic-intro}) and introduce two improvements.  The first one
concerns the inclusion of realistic frequency uncertainties, to compute
reduced $\chi^2$ values as a means of the model selection criterion.  
P14 only adopted a first
rough goodness-of-fit procedure for the model selection, denoted here as
$\chi^2_{\rm P14}$. This goodness-of-fit was based on the conservative Rayleigh
limit of 0.000685\,d$^{-1}$ for the frequency error of all 19 detected modes,
with the argument that the uncertainties on theoretical predictions of
oscillation frequencies of pulsating B stars are typically about
0.001\,d$^{-1}$ \citep{briquet-2007-01}. That was adequate for their modelling,
because they only compared models with one set of input physics and with one fixed
number of degrees of freedom.  Since we here compare
models with and without extra diffusive mixing, and thus with a different number
of degrees of freedom, we must use a more appropriate statistical description to
perform the model selection. This requires the determination of individual
frequency uncertainties, as explained in Sect.\,\ref{s-kic-intro}. 
The second improvement concerns a more detailed model comparison, where we
consider different elements of the input physics, such as opacities,
chemical mixtures and two different descriptions for the core
overshooting. Moreover, we add models based on the inclusion of extra diffusive
mixing outside the convective and overshooting zones, described by a coefficient
$D_{\rm mix}$.

We introduce the target star in Sect.\,\ref{s-kic-intro} and explain
how we took into account its seismic data in Sect.\,\ref{s-dP}.
The physical ingredients of  the evolutionary models computed with the 
MESA code 
\citep[][version 5548]{paxton-2011-01, paxton-2013-01} are explained 
in Sect.\,\ref{s-kic-grid}.
The adiabatic zonal 
dipole frequencies for each input model along the evolutionary tracks 
were computed with the GYRE pulsation code 
\citep[][version 3.0]{townsend-2013-01}.
In Sect.\,\ref{s-chisq}, we explain the model selection based on a
reduced $\chi^2$ method.
The results of our grid search to select 
the best models to explain the seismic data  of KIC\,10526294 are reported in 
Sect.\,\ref{s-chisq}.
Our consideration of three metal mixtures and both OP and 
OPAL opacities is exploited in terms of 
the mode stability for the best models in 
Sect.\,\ref{s-excitation}. 
Finally, we discuss our results in Sect.\,\ref{s-discussion}.


\section{The showcase of KIC\,10526294}
\label{s-kic-intro}

P14 identified KIC\,10526294 as a new SPB star \citep[e.g.][for a
definition]{waelkens-1991-01, de-cat-2002-01} from its {\it Kepler\/} light
curve, which has a duration of $\Delta T$ = 1460\,d.  Its effective temperature,
surface gravity, and metallicity were determined by P14 from high-resolution
spectroscopy: $T_{\rm eff}=11\,500\pm500$ K, $\log g=4.1\pm0.2$ dex, and
$Z=0.016_{-0.007}^{+0.013}$.  P14 performed a full study of the
\textit{Kepler\/} photometry and identified 19 consecutive dipole $\ell=1$
g-mode triplets.  As mentioned above, we consider these 19 zonal-mode
frequencies with their appropriate errors. The latter were determined following
the methodology discussed in detail in \citet{degroote-2009-01}, which is based
on the theory of time-series analysis of correlated data by
\citet{schwarzenberg-czerny-1991-01}. In practice this implies taking the
formal errors of the non-linear least-squares fit to the light curve and to
correct them for the signal-to-noise ratio, sampling, and correlated nature of
the data. Applying the procedure discussed in \citet{degroote-2009-01} to
KIC\,10526294 results in a correction factor of 3.0 to be applied to the formal
errors listed in Table\,B.1 in P14.  For convenience, we repeat the 19 zonal
dipole frequencies, their periods and period spacings and their corresponding 
1$\sigma$ errors in Table\,\ref{t-freq}.

As discussed in P14, the average rotational frequency splitting for the $m=\pm1$
components of the dipole modes points to a rotation period of $\sim188$\,d when
averaged over the entire depth of the star, implying that KIC\,10526294 is an
ultra-slow rotator.  According to the survey of galactic B stars by
\citet{huang-2010-01}, the probability of finding a very slowly rotating late B-type
star (2 $\leq$ M/M$_\odot$ $\leq$ 4) is low (their Fig.\,7a).  Thus, KIC\,10526294 is optimally
suited to study mixing mechanisms for a case where the effect of rotation is
negligible.

Based on the initial asteroseismic modelling presented in P14, the mass of 
KIC\,10526294
is roughly 3.2\,M$_\odot$, and the star is situated very close to the
ZAMS.  P14 assumed convective core overshooting and semi-convective mixing 
\citep[according to the prescription by][with $\alpha_{\rm sc}=10^{-2}$]{
langer-1983-01}.
They found their best seismic models (their Table\,6) to have a
core overshooting $f_{\rm ov}$ in the range from zero to 0.015.  In what
follows, we start from the same grid of models as in P14 but consider additional
extra mixing, various chemical mixtures, different opacity tables and also a
step function for the core overshooting. For all these cases, we study how 
the goodness-of-fit behaves and perform model comparisons.

\begin{table}
\begin{center}
\caption{Frequencies $f_i^{\rm(obs)}$, periods $P_i^{\rm (obs)}$, 
period spacings $\Delta P_i^{\rm(obs)}$, and 1$\sigma$ errors of the central 
peaks ($m=0$) of the dipole ($\ell=1$) triplets observed in KIC\,10526294
as determined from Table\,B.1 in P14.} 
\label{t-freq}
\tabcolsep=4pt
\begin{tabular}{cccc}
  \hline
  ID       & $f_i^{\rm(obs)}\pm\sigma_i$ & $P_i^{\rm (obs)}\pm\delta P_i$ & 
$\Delta P_i^{\rm(obs)}\pm\epsilon_i$ \\           & [d$^{-1}$] & [d]    & [s] \\
  \hline
   $f_{1}$  & 0.472220$\pm$0.000057 & 2.11766$\pm$0.000256 & --- \\
   $f_{2}$  & 0.486192$\pm$0.000057 & 2.05680$\pm$0.000241 & 5258$\pm$30 \\
   $f_{3}$  & 0.500926$\pm$0.000036 & 1.99630$\pm$0.000143 & 5227$\pm$24 \\
   $f_{4}$  & 0.517303$\pm$0.000051 & 1.93310$\pm$0.000191 & 5460$\pm$21 \\
   $f_{5}$  & 0.533426$\pm$0.000018 & 1.87467$\pm$0.000063 & 5048$\pm$17 \\
   $f_{6}$  & 0.552608$\pm$0.000006 & 1.80960$\pm$0.000020 & 5622$\pm$6 \\
   $f_{7}$  & 0.571964$\pm$0.000012 & 1.74836$\pm$0.000037 & 5291$\pm$4 \\
   $f_{8}$  & 0.593598$\pm$0.000039 & 1.68464$\pm$0.000111 & 5505$\pm$10 \\
   $f_{9}$  & 0.615472$\pm$0.000018 & 1.62477$\pm$0.000048 & 5173$\pm$10 \\
   $f_{10}$ & 0.641202$\pm$0.000069 & 1.55957$\pm$0.000168 & 5633$\pm$15 \\
   $f_{11}$ & 0.670600$\pm$0.000057 & 1.49120$\pm$0.000127 & 5907$\pm$18 \\
   $f_{12}$ & 0.701246$\pm$0.000075 & 1.42603$\pm$0.000153 & 5631$\pm$17 \\
   $f_{13}$ & 0.734708$\pm$0.000009 & 1.36108$\pm$0.000017 & 5612$\pm$13 \\
   $f_{14}$ & 0.772399$\pm$0.000045 & 1.29467$\pm$0.000075 & 5738$\pm$7 \\
   $f_{15}$ & 0.812940$\pm$0.000009 & 1.23010$\pm$0.000014 & 5578$\pm$7 \\
   $f_{16}$ & 0.856351$\pm$0.000030 & 1.16775$\pm$0.000041 & 5388$\pm$4 \\
   $f_{17}$ & 0.902834$\pm$0.000069 & 1.10762$\pm$0.000085 & 5195$\pm$8 \\
   $f_{18}$ & 0.954107$\pm$0.000030 & 1.04810$\pm$0.000033 & 5143$\pm$8 \\
   $f_{19}$ & 1.013415$\pm$0.000036 & 0.98676$\pm$0.000035 & 5300$\pm$4 \\
  \hline
\end{tabular}
\end{center}
\end{table}

\section{Period spacings of the gravity modes}
\label{s-dP}

The probing power of asteroseismology of SPB stars emerges from (a) the
sensitivity of the g-mode frequencies to the physics of the stellar interior in
general and to the propagation cavities in particular, (b) the relative
eigenfunction variation across the star, and (c) mode trapping in the overshooting
zone \citep[e.g.,][]{dziembowski-1991-01, miglio-2008-01}.
At a fixed age (associated with the hydrogen mass fraction in the core $X_{\rm c}$),
some modes have a node near the chemically inhomogeneous zones and are able to
probe the overshooting layer.

\begin{figure}
\includegraphics[width=\columnwidth]{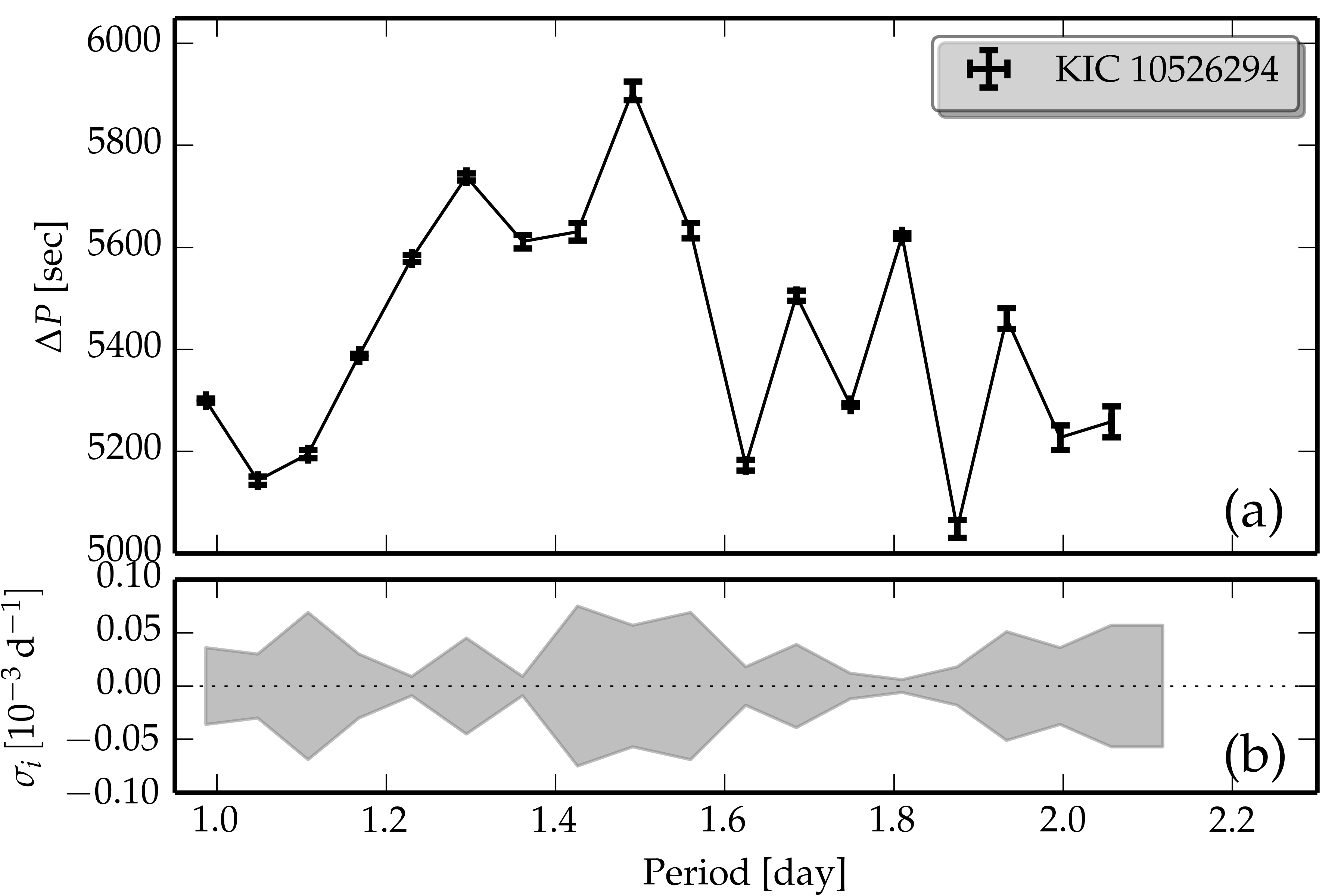}
\caption{Top. The period spacing $\Delta P$ for the 19 dipole gravity modes
observed in the \textit{Kepler\/} SPB star KIC\,10526294.
Bottom. The 1$\sigma$ uncertainty region around each frequency $f^{\rm(obs)}_i$
as given in Table\,\protect\ref{t-freq}.}
\label{f-obs-dP}
\end{figure}

The asymptotic period spacing for high-order low-degree gravity modes 
for a non-rotating star is given by
\begin{equation}\label{e-dP}
\Delta P_{n,\ell} = P_{n+1,\ell}-P_{n,\ell} = \frac{2\pi^2}{\sqrt{\ell(\ell+1)}}
\left(\int_{R_{\rm cc}}^{R_\star}\frac{N(r)}{r}\,dr\right)^{-1},
\end{equation}
where $R_\star$ is the stellar radius, $R_{\rm cc}$ indicates the boundary
of the  convective core, 
and $P_{n,\ell}$ and $P_{n+1,\ell}$ are periods of two consecutive modes
in radial order of the same degree $\ell$ \citep{tassoul-1980-01}.
In practice, the spatial integration is only carried out in the region
  where the g-modes propagate.
Since we only consider consecutive dipole modes, we drop the $n$ and $\ell$
subscripts from $\Delta P_{n,\,\ell}$ in the rest of the paper.

Fig.\,\ref{f-obs-dP}a presents the sequence of 18 period spacings for the 
19 consecutive  dipole g-modes detected in KIC\,10526294.
They exhibit a small but clear deviation from the asymptotic (i.e.\ constant) 
period spacing relation in Eq.\,(\ref{e-dP}) 
and highlight the presence of chemical inhomogeneities that are 
left behind by the shrinking convective core \citep{miglio-2008-01}.
In Fig.\,\ref{f-obs-dP}b, the frequency uncertainties $\sigma_i$ of
each frequency $f^{\rm(obs)}_i$ are presented as a function of the mode periods. 

\begin{figure}
\includegraphics[width=\columnwidth]{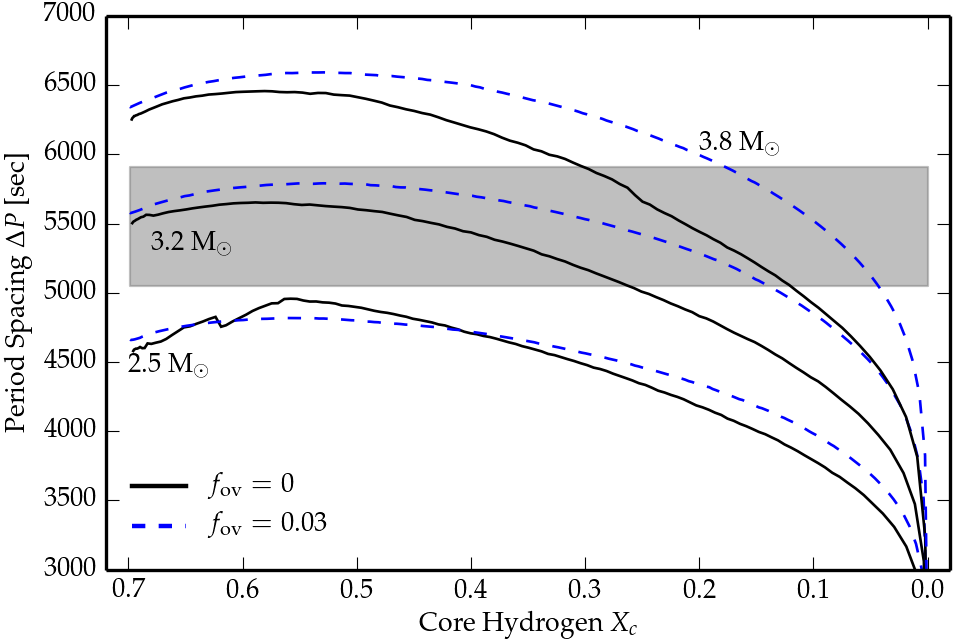}
\caption{Evolution of the asymptotic period spacing for dipole modes
computed from the  right-hand side of  Eq.\,(\ref{e-dP}) 
versus $X_{\rm c}$ for models with masses 2.5 M$_\odot$, 3.2 M$_\odot$, 
and 3.8 M$_\odot$, each for $Z_{\rm ini}=0.020$.  
The solid lines correspond to models without overshooting, while the dashed lines 
correspond to models including core overshooting with a value of $f_{\rm ov}=0.03$.  
The thin grey bar highlights the measured period range of 5048 to 5907\,s for 
KIC\,10526294 from Table\,\ref{t-freq}; see also Fig.\,\ref{f-obs-dP}.
All models were computed using OPAL opacity tables, and the 
\citet{nieva-2012-01} and 
\citet{przybilla-2013-01} mixture.}
\label{f-dP-evol-Xc}
\end{figure}

During the MS phase, the core gradually shrinks and its density and pressure increase.
Thus, the integrand in the right-hand side of Eq.\,(\ref{e-dP}) increases, and the asymptotic value  
$\Delta P$ as well as the structure of the period spacing sequence evolve.
The former depends on the size of the fully mixed core, while the latter depends
on the detailed treatment of  mixing in the radiative envelope on top of the shrinking 
core \citep{miglio-2008-01}.
Fig.\,\ref{f-dP-evol-Xc} shows the evolution of the asymptotic $\Delta P$
value computed from the right-hand side of 
Eq.\,(\ref{e-dP}) for three models with masses of 2.5\,M$_\odot$, 
3.2\,M$_\odot$, and 3.8\,M$_\odot$.
The observed period spacing range for KIC\,10526294 from Table\,\ref{t-freq} is 
highlighted in grey.  
In practice, we chose the median frequency, $f$, from Table\,\ref{t-freq}
and carried out the integration in the region 
where $f^2\leq N^2\leq S_{\ell}^2$ (with $S_{\ell}$ the Lamb frequency). 
Because the frequency range of $f_1$ to $f_{19}$ is limited, the choice of $f$ does not 
alter the result of the integration, as we have verified.
The asymptotic period spacing for models without overshooting 
($f_{\rm ov}=0.00$, solid black lines) are mostly below those for models 
with overshooting ($f_{\rm ov}=0.03$, blue dashed lines).
Varying the initial metallicity $Z_{\rm ini}$ introduces similar changes, 
which are not shown here for clarity.

Because several combinations of the stellar parameters can reproduce the same
asymptotic $\Delta P$ value, it is essential to fit individual frequencies 
in addition to matching the asymptotic $\Delta P$ when performing seismic modelling.
Based only on Fig.\,\ref{f-dP-evol-Xc}, both models at the low and high side of
the mass range $[3,4]$\,M$_\odot$ are able to explain the observed 
$\Delta P$ for KIC\,10526294. Additional tuning and model selection 
requires scanning dense asteroseismic model grids that have a sufficiently broad range
in mass $M$, but are also sufficiently fine in terms of all the other input parameters 
$X_{\rm ini}$, $Z_{\rm ini}$, $X_{\rm c}$, $\log D_{\rm mix}$, and 
$f_{\rm ov}$, to obtain a meaningful result, according to the precision of
the identified frequencies. 

\section{Extensive evolutionary and asteroseismic grids}
\label{s-kic-grid}

P14 presented a dense evolutionary and asteroseismic grid of models dedicated to
KIC\,10526294; they considered relatively young models ($X_{\rm c}\geq0.40$) in the
range of 3.00 to 3.40 M$_\odot$.
Here, we followed the same approach, but we also included 
extra diffusive mixing
beyond the convective and overshooting zones.
Moreover, we varied several input parameters 
to study if that improves the quality of the frequency fitting compared to the 
one achieved by P14.
We present eight evolutionary and asteroseismic grids in which some parameters
are varied and some are kept fixed to keep the CPU requirements manageable.

The physical ingredients of the grids are the following.
We used the open-source MESA\footnote{MESA can be downloaded from the project Web site 
\href{http://mesa.sourceforge.net}{http://mesa.sourceforge.net}} 
code \citep{paxton-2011-01, paxton-2013-01} to
calculate evolutionary tracks and store equilibrium models along each track for
every 0.001 drop in $X_{\rm c}$.
This tiny grid step in the central hydrogen fraction was taken to
  ensure a small enough change in the frequency values of the modes as we follow
  the stellar evolution along an evolutionary track, relying on
  the numerical accuracy achieved by the MESA code coupled to the pulsation code we
  used (as discussed below), cf.\ Fig.\,13 in P14 and its discussion.

We adopted the Ledoux convection criterion with a high semi-convective mixing
coefficient given by $\alpha_{\rm sc}=10^{-2}$ in the prescription by
\citet{langer-1983-01}. 
For this value, the Ledoux criterion is equivalent to the Schwarzschild criterion
and ensures that $\nabla_{\rm rad}=\nabla_{\rm ad}$ on the convective side of
the core, following \citet{gabriel-2014-01}.
The mixing length parameter was set to $\alpha_{\rm MLT}=1.8$, while we fixed the
initial composition to the Galactic standard given by \citet{nieva-2012-01} and
\citet[][hereafter NP12]{przybilla-2013-01}: 
$(X_{\rm ini},Y_{\rm ini},Z_{\rm ini})=(0.710, 0.276, 0.014)$.
We used the OPAL Type 1 \citep{rogers-2002-01} opacity tables adapted to this
mixture; these tables and the MESA inlists are available for download as online
material (see Appendix\,\ref{s-deliv}).

We adopted an exponentially decaying prescription for overshooting on top of the
convective core in a diffusive approximation, following \citet{freytag-1996-01}
and \citet{herwig-2000-01}:
\begin{equation}\label{e-exp-ov}
D_{\rm ov}(z) = D_{\rm conv} \, \exp{\left(-\frac{2 \, z}{f_{\rm ov}\, H_p}\right)},
\end{equation}
where $D_{\rm conv}$ is the convective mixing coefficient, 
$z=r-R_{\rm cc}$ is the distance
above the boundary of the convective core $R_{\rm cc}$ into the radiative
region, and 
$H_p$ is the local pressure scale height.
The parameter $f_{\rm ov}$ governs the width of the overshooting layer,
which in the framework of the mixing length theory of \cite{bohm-vitense-1958-01},
is not constrained from first principles.
We therefore treated it here as a free parameter and 
varied it in a broad range from 0.0 to 0.03 in all our grids.

\begin{figure*}
\includegraphics[width=\textwidth]{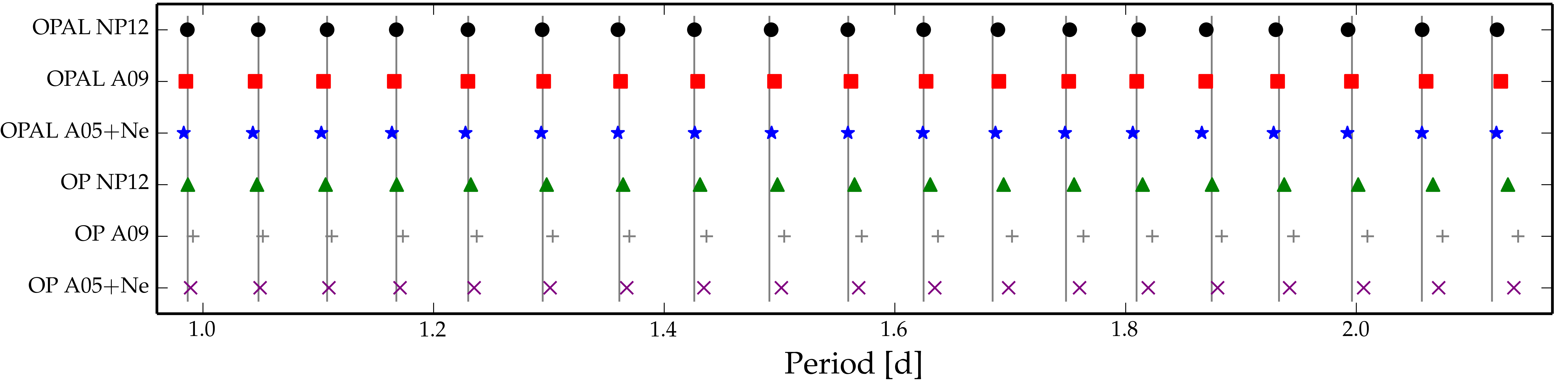}
\caption{Comparison between observed periods (vertical lines) and those of six 
models with different mixtures and opacity tables.
All parameters of the models are identical, except their mixture and opacities.
The widths of the vertical lines are larger than the measured pulsation 
period uncertainties.}
\label{f-mixture-freq-comp}
\end{figure*}

An alternative for core overshooting is to use a step-function prescription, which is 
commonly used in the literature.
In this prescription, the overshooting extends over a distance 
$d_{\rm ov}=\alpha_{\rm ov} H_p$, 
beyond the convective core boundary. 
The material is treated as fully mixed in the overshoot range:  
\begin{equation}
\label{e-step-ov}
   D_{\rm ov}=D_{\rm conv}, \quad R_{\rm cc} \leq r \leq R_{\rm cc}+d_{\rm ov},
\end{equation}
where $R_{\rm cc}$ is the radial position of the boundary of the convective core,
$d_{\rm ov}$ is the width of the overshooting zone and $D_{\rm conv}$ is the diffusion 
coefficient in the fully mixed convective core.
Following the MESA implementation of core overshooting according to a step
function, we tested explicitly that our seismic models 
are insensitive to the depth in the convective core at which the value of 
$D_{\rm conv}$ is picked up.

Our eight extensive evolutionary and asteroseismic grids are composed and 
named as follows:
\begin{itemize}
\item \textbf{Basic Grid:} We investigated whether more massive and more evolved
  models may reproduce the observed, nearly flat, period spacing pattern of
  KIC\,10526294 as well.  The Basic Grid is essentially the extension of the
  grid in P14 towards higher masses and lower $X_{\rm c}$.

\item \textbf{Composition Grid:} We calculated a grid varying $X_{\rm ini}$ and
  $Z_{\rm ini}$ for the NP12 mixture to investigate if this improves the
  frequency fit.

\item \textbf{Mixing Grid:} Various extra mixing mechanisms can operate and lead
  to smoothing of the composition profile outside the core.  Unfortunately, we
  have no solid quantitative measures of the amount of extra mixing for B stars.
  Hence we computed a grid with models
  including extra diffusive mixing throughout the star for a wide range of
  $\log\,D_{\rm mix}$ values.

\item \textbf{Evolved Grid:} A large amount of extra mixing combined with a low
  $X_{\rm c}$ may provide a situation for which the $\Delta P$ profile matches
  the observations for higher mass models. 
  In Fig.\,\ref{f-dP-evol-Xc}, higher mass models cross the grey band at lower
  $X_{\rm c}$.  To ensure our age constraint is reliable, we calculated a grid
  for $M\geq3.30$\, M$_\odot$ incorporating extra mixing; we only considered the
  models whose mean $\Delta P$ falls inside the observed range in
  Fig.\,\ref{f-dP-evol-Xc}.  Thus, the starting and ending $X_{\rm c}$ values
  for each track are flexible. This is the meaning of ``---'' in
  Table\,\ref{t-grid}.

\item \textbf{Fine Grid:}
  To resolve the parameter space of mass, overshooting, and extra mixing, 
  we increased the resolution for these parameters
  around the best model found from the Mixing Grid.
  Thus, we varied $M \in[3.18, \,3.27]$ M$_\odot$ in steps of 0.01 M$_\odot$, 
  $f_{\rm ov} \in[0.016, 0.019]$ in steps of 0.001, and 
  $\log D_{\rm mix}\in[0.25, \,2.75]$ in steps of 0.25 dex.

\item \textbf{Metallicity Grid:} Although the average metallicity of
    Galactic B stars in the solar neighbourhood is found to be $Z=0.014\pm0.002$
    \citep{nieva-2012-01}, individual B stars may still show a strong deviation
    from this average.  Enhanced metallicity increases the height of the iron
    opacity bump and ensures better mode excitation. Moreover, as we demonstrate
    below in Fig.\,\ref{f-corr-planes}, the initial metallicity and mass exhibit
    a strong anti-correlation.  To explore the possibility of our target having
    much higher metal abundance than the cosmic standard, we computed a grid
    identical to the Mixing Grid and Fine Grid, but with lower initial mass and
    higher initial metallicity up to $Z_{\rm ini}=0.030$. 

\item \textbf{Mixture Grids:} The theoretical pulsation frequencies depend on
  the initial metal mixture and the adopted opacity tables.  To illustrate this,
  Fig.\,\ref{f-mixture-freq-comp} compares the observed periods (vertical grey
  lines) with those of six models with identical model parameters, except that
  they have different mixtures and opacities.  The adopted mixtures for the grid
  are NP12, the solar mixture reported by \citet{asplund-2009-01} (hereafter A09) and the
  \citet{asplund-2005-01} mixture with a Ne enhancement based on
  \citet{cunha-2006-01} (hereafter A05$+$Ne).  Fig.\,\ref{f-mixture-freq-comp}
  shows a clear difference in the frequency predictions, particularly for the
  longer-period modes.  The relative difference in periods ranges
  from 0.07\% to 0.14\% from left to right in the figure.  Changing the
  mixture in the models influences fitting high-precision frequencies of
  well-identified pulsation modes.  In this respect, the asteroseismic data of
  KIC\,10526294 bring us to the level where the dependence of the models on the
  assumed mixture can be tested for SPB stars, thanks to the identification of
  its dipole mode triplets.

  We calculated grids using both OP \citep{seaton-1987-01} and OPAL
  \citep{rogers-2002-01} opacity tables for each of the NP12, A09, and A05$+$Ne
  mixtures.  The parameter range for each of these sub-grids is identical.  Once
  we found the best model in each sub-grid, we repeated the computations within a
  smaller range around the best model until the step size in mass, overshooting,
  metallicity, and $\log D_{\rm mix}$ was 0.01, 0.001, 0.001 and 0.25 dex,
  respectively.

\item \textbf{Step Function Overshoot Grid:} To examine which of the two
  prescriptions for the core overshooting, exponentially decaying as in
  Eq.\,(\ref{e-exp-ov}) versus step function as in Eq.\,(\ref{e-step-ov}),
  provides a better fit to the observed data, a stand-alone grid of models for a
  step function overshoot was computed around the best model from the Fine
  Grid, with the only change of using the step function instead of the
  exponential prescription for a range of masses and metallicities.
\end{itemize}

Table\,\ref{t-grid} gives the full parameter range in each of these grids.
In the Mixing Grid and High Mass Grid, we varied the minimum diffusive mixing
coefficient $D_{\rm mix}$.  For the Basic Grid and Composition Grid, we set
$D_{\rm mix}=0$.  In all grids except the Step Overshoot Grid, we used the
exponential prescription as in Eq.\,(\ref{e-exp-ov}).  In total, our grids
comprise 27\,058 evolutionary tracks and around 4.3 million 
models that serve as input for the GYRE pulsation computations.

\begin{table}[h!]
\centering
\caption{Eight model grids and their parameter range dedicated to the
  asteroseismic modelling of KIC\,10526294.
$N$ is the number of values for each parameter.
$\alpha_{\rm sc}$ is fixed to $10^{-2}$ in all models.
$D_{\rm mix}$ is in cm$^2$\,s$^{-1}$.
$n$ is the number of degrees of freedom in Eq.\,(\ref{e-chisq-freq}).
For explanations of $X_{\rm c}$ in the Evolved Grid, we refer to the text.
The Mixture Grid is repeated for OP and OPAL opacity tables and for NP12, A09 and 
A05$+$Ne mixtures.} 
\label{t-grid}
{\small
\begin{tabular}{lllll}
  \hline
  Parameter   & From & To     & Step & $N$ \\
  \hline
  \textit{Basic} Grid ($n$=4): \\
  Ini. mass:  M$_{\rm ini}$ [M$_\odot$] & 3.00  & 3.95  & 0.05 & 25 \\
  Overshooting:  $f_{\rm ov}$ & 0.000 & 0.030 & 0.003 & 11 \\
  Ini. hydrogen:   $X_{\rm ini}$ & 0.71 & 0.71 & 0.01 & 1 \\
  Ini. metallicity:   $Z_{\rm ini}$ & 0.010 & 0.020 & 0.001 & 11 \\
  Centre hydrogen:  $X_{\rm c}$ & 0.700 & 0.0 & 0.001 & $>$701	\\
  \\[-6pt]
  \textit{Composition} Grid ($n$=5): \\
  Ini. mass:  M$_{\rm ini}$ [M$_\odot$] & 3.15  & 3.25  & 0.05 & 3 \\
  Overshooting:  $f_{\rm ov}$ & 0.000 & 0.030 & 0.003 & 11 \\
  Ini. hydrogen:   $X_{\rm ini}$ & 0.68 & 0.72 & 0.01 & 5 \\
  Ini. metallicity:   $Z_{\rm ini}$ & 0.010 & 0.020 & 0.001 & 11 \\
  Centre hydrogen:  $X_{\rm c}$ & 0.700 & 0.600 & 0.001 & 101 \\
  \\[-6pt]
  \textit{Mixing} Grid ($n$=5): \\
  Ini. mass:  M$_{\rm ini}$ [M$_\odot$] & 3.10  & 3.30  & 0.05 & 5 \\
  Overshooting:  $f_{\rm ov}$ & 0.000 & 0.030 & 0.003 & 11 \\
  Ini. hydrogen:   $X_{\rm ini}$ & 0.71 & 0.71 & 0.01 & 1 \\
  Ini. metallicity:   $Z_{\rm ini}$ & 0.010 & 0.020 & 0.001 & 11 \\
  Centre hydrogen:  $X_{\rm c}$ & 0.700 & 0.600 & 0.001 & 101 \\
  Extra Mixing: $\log D_{\rm mix}$ & 2 & 6 & 1 & 5 \\
  \\[-6pt]
  \textit{Evolved} Grid ($n$=5): \\
  Ini. mass:  M$_{\rm ini}$ [M$_\odot$] & 3.30  & 4.00  & 0.05 & 15 \\
  Overshooting:  $f_{\rm ov}$ & 0.000 & 0.030 & 0.006 & 6 \\
  Ini. hydrogen:   $X_{\rm ini}$ & 0.71 & 0.71 & 0.01 & 1 \\
  Ini. metallicity:   $Z_{\rm ini}$ & 0.010 & 0.020 & 0.002 & 6 \\
  Centre hydrogen:  $X_{\rm c}$ & --- & --- & 0.001 & --- \\
  Extra Mixing: $\log D_{\rm mix}$ & 2 & 6 & 1 & 5 \\
  \\[-6pt]
  \textit{Fine} Grid ($n$=5): \\
  Ini. mass:  M$_{\rm ini}$ [M$_\odot$] & 3.18  & 3.27  & 0.01 & 10 \\
  Overshooting:  $f_{\rm ov}$ & 0.000 & 0.030 & 0.003 & 11 \\
                              & 0.016 & 0.019 & 0.001 & 3 \\
  Ini. hydrogen:   $X_{\rm ini}$ & 0.71 & 0.71 & 0.01 & 1 \\
  Ini. metallicity:   $Z_{\rm ini}$ & 0.010 & 0.020 & 0.001 & 11 \\
  Centre hydrogen:  $X_{\rm c}$ & 0.700 & 0.600 & 0.001 & 101 \\
  Extra Mixing: $\log D_{\rm mix}$ & 0.25 & 2.75 & 0.25 & 10 \\
  \\[-6pt]
  \textit{metallicity} Grid ($n$=5): \\
  Ini. mass:  M$_{\rm ini}$ [M$_\odot$] & 3.00  & 3.20  & 0.05 & 5 \\
                                        & 3.12  & 3.18  & 0.01 & 7 \\
  Overshooting:  $f_{\rm ov}$ & 0.012 & 0.024 & 0.003 & 5 \\
  Ini. hydrogen:   $X_{\rm ini}$ & 0.71 & 0.71 & 0.01 & 1 \\
  Ini. metallicity:   $Z_{\rm ini}$ & 0.021 & 0.030 & 0.001 & 11 \\
  Centre hydrogen:  $X_{\rm c}$ & 0.700 & 0.600 & 0.001 & 101 \\
  Extra Mixing: $\log D_{\rm mix}$ & 1.00 & 2.00 & 0.25 & 5 \\
  \\[-6pt]
  \textit{Mixture} Grids ($n$=5): \\
  Ini. mass:  M$_{\rm ini}$ [M$_\odot$] & 3.17  & 3.29  & 0.02 & 7 \\
  Overshooting:  $f_{\rm ov}$ & 0.012 & 0.024 & 0.003 & 5 \\
  Ini. hydrogen:   $X_{\rm ini}$ & 0.71 & 0.71 & 0.01 & 1 \\
  Ini. metallicity:   $Z_{\rm ini}$ & 0.010 & 0.018 & 0.002 & 5 \\
  Centre hydrogen:  $X_{\rm c}$ & 0.700 & 0.600 & 0.001 & 101 \\
  Extra Mixing: $\log D_{\rm mix}$ & 1.50 & 2.50 & 0.25 & 5 \\
  \\[-6pt]
  \textit{Step Overshoot} Grid ($n$=5): \\
  Ini. mass:  M$_{\rm ini}$ [M$_\odot$] & 3.18  & 3.27  & 0.01 & 10 \\
  Overshooting:  $\alpha_{\rm ov}$ & 0.16 & 0.23 & 0.01 & 8 \\
  Ini. hydrogen:   $X_{\rm ini}$ & 0.71 & 0.71 & 0.01 & 1 \\
  Ini. metallicity:   $Z_{\rm ini}$ & 0.010 & 0.020 & 0.001 & 10 \\
  Centre hydrogen:  $X_{\rm c}$ & 0.700 & 0.600 & 0.001 & 101 \\
  Extra Mixing: $\log D_{\rm mix}$ & 1.50 & 2.25 & 0.25 & 4 \\
  \hline
\end{tabular}}
\end{table}

For each input model, we used the state-of-the-art GYRE code
\citep[][version 3.0]{townsend-2013-01} to solve for the linear adiabatic oscillation
frequencies of dipole $\ell=1$, $m=0$ modes in the frequency range of the
detected g-modes of KIC\,10526294.
The input inlist for the GYRE computations 
is available for download as online material; 
see Appendix\,\ref{s-deliv}.

\section{Model selection}\label{s-chisq}

\subsection{Methodology}\label{ss-chisq}

We considered a reduced $\chi^2_{\rm red}$ minimisation scheme to
select the best-fitting models after comparing the observed frequencies
$f_i^{\rm (obs)}$ with their 
model counterparts $f_i^{\rm (th)}$:
\begin{align}\label{e-chisq-freq} 
 \chi^2_{\rm red} &= \frac{1}{N_f-n} \sum_{i=1}^{N_f}\frac{(f^{\rm (obs)}_i -
   f^{\rm (th)}_i)^2}{\sigma_{i}^2}, 
 \end{align}
 where $N_f=19$ is the number of consecutive g-modes and $\sigma_i$ are the
 theoretical frequency uncertainties listed in Table\,\ref{t-freq}.  The number
 of degrees of freedom, $n$ in Eq.\,(\ref{e-chisq-freq}), represents the number
 of independent model parameters.  In all the cases, $n$ is either 4 or 5, as
 listed in Table\,\ref{t-grid}.  
 This procedure is a simplified
version of the goodness-of-fit adopted by P14, who also included the period spacing
values in the $\chi^2$.  
Here, we prefered to forego this because demanding an optimal fit to the observed
frequencies individually will automatically also deliver an optimal fit to the
observed period spacing values. Moreover, we wished to compare the fit quality of
models with a different number of degrees of freedom ($n=4$ or 5, cf.\
Table\,\ref{t-grid}) for the best model selection and this is better justified
statistically when using independent measurements in the computation of 
$\chi^2_{\rm  red}$.

The $\chi^2_{\rm red}$ values in Eq.\,(\ref{e-chisq-freq}) gauge how well the
model frequencies match the observed ones, where a value of 1 would mean that
the input physics of the models is correct up to the level of the frequency
precision.  We are very far from this situation in the case of stars of
spectral type O, B, or A. Indeed, for such massive stars, we are in a different
situation compared to solar-like stars with stochastically excited oscillations,
for which scaling relations can be applied to the {\it Kepler\/} data 
\citep{chaplin-2014-01} and
where extrapolations from the solar model are meaningful. In these straightforward cases, 
one achieves seismic modelling with typical values of $\chi^2_{\rm
  red}\sim500$ for the best {\it Kepler\/} data of G- or F-type stars on the
main sequence, is achieved with a grid-based approach like the one we adopt here
\citep{metcalfe-2014-01}.   
For the best modelled subgiants with solar-like oscillations, 
$\chi^2_{\rm red}$-values between 100 to 2000 were obtained 
\citep[e.g.][]{deheuvels-2014-01}.

Seismic modelling of massive stars with
heat-driven modes has not been done so far using the level of precision of the
measured frequencies, cf.\ the discussion in Sect.\,\ref{s-intro}
concerning the use of the Rayleigh limit as rough estimate of the measured frequency
precision. Theoretical frequency predictions from models have typically only been
computed up to 0.001\,d$^{-1}$, and the models were selected by
visual inspection of the frequency spectra \citep[e.g.][]{pamyatnykh-2004-01} or by a
$\chi^2$-type goodness-of-fit function ignoring the measured frequency
precisions \citep[e.g.][]{briquet-2007-01}. This is appropriate as a procedure when
  only a handful of identified mode frequencies is available.
In this way, stellar masses, radii, core overshooting values, 
and ages have been deduced \citep[e.g.][for an overview]{aerts-2015}. 

\subsection{Candidate models}\label{ss-candidates}

Table\,\ref{t-chisq} lists the parameters of
the best-fit models for each of the grids described in Table\,\ref{t-grid}.  The
first and second columns assign a number to each model and specify their grid of
origin, respectively.  
The next eight columns specify the adopted input
parameters for the best models.  The last two columns give the $\chi^2_{\rm red}$
(Eq.\,\ref{e-chisq-freq}) and $\chi^2_{\rm P14}$ values, respectively. 
The latter column is given to provide a sensible comparison with the results obtained 
from our previous modelling as discussed in P14 (their Table\,6, Col.\,11), 
where the input physics was kept fixed.

Model\,1 is essentially the replication of the best model found in P14 from the
Basic Grid, as a verification of the consistency between our previous and current study.
Model\,2 is taken from the Composition Grid.
Comparison between Model\,1 and Model\,2 shows that, for the specific case of 
KIC\,10526294, changing the initial composition $X_{\rm ini}$ and $Y_{\rm ini}$ 
deteriorates the frequency fittings.
Thus, the Galactic standard composition of B stars by \citet{nieva-2012-01} is the 
most appropriate one to take for this target.
Model\,3 is the best of its kind from the Evolved Grid, but its $\chi^2_{\rm red}$ 
is significantly higher than that of Model\,1.
Therefore, we exclude the possibility that KIC\,10526294 is an evolved star.

Model\,4 is taken from the Mixing Grid and the Fine Grid.  It has the lowest
$\chi^2_{\rm red}$ score and is the best seismic model for KIC\,10526294.  
It outperforms all models without diffusive mixing and is roughly 11 times better 
in fit quality than Model\,1.
Thus, we conclude that the frequency fitting for
KIC\,10526294 requires extra diffusive mixing with a value of $\log D_{\rm
mix}=1.75$ cm$^2$\,s$^{-1}$. 
This is a similar conclusion as was reached for HD\,50230
\citep{degroote-2010-01}, only this time we were able to quantify $D_{\rm mix}$
much more precisely and we considered various options for the metal mixture and
opacities, while $Z=0.02$ was fixed for HD\,50230 and the degrees of its
  modes were not identified observationally.
We find a value for the diffusive mixing
coefficient of KIC\,10526294 that is two orders of magnitude lower
than the one for HD\,50230.
The two major differences between these stars are their mass and
their $X_c$, both being ultra-slow rotators. It would be worth revisiting the
seismic modelling of HD\,50230 now that we are aware that it is a member of a
spectroscopic binary and that p-modes have been found in addition to g-modes
\citep{degroote-2012-01}; these two facts were not taken into
account in the modelling efforts made by \citet{degroote-2010-01}.

\begin{table*}
\begin{center}
  \caption{Compilation of best models from different grids (Table\,\ref{t-grid})
    and their input parameters.  These models have the lowest $\chi^2_{\rm
      red}$ scores within each grid.  The overshooting parameter
    is $f_{\rm ov}$ for all models in Col.\,9, except for Model\,10 where we list
    $\alpha_{\rm ov}$.  Models\,4, 8 and 11 represent the
    best three seismic models of KIC\,10526294; for each of them, the
      radial orders of the 19 detected modes range from 14 to 32.  The overall
    best model in a statistical sense is indicated in bold.}
\label{t-chisq}
\begin{tabular}{llllllllllll}
  \hline
  \#    & Grid & Opacity & Mixture & Mass & $X_{\rm ini}$ & $Z_{\rm ini}$ & $X_{\rm c}$ & 
  $f_{\rm ov}$ & $\log D_{\rm mix}$ & $\chi^2_{\rm red}$ & $\chi^2_{\rm P14}$\\
        &  &  &  & [M$_\odot$] &  &  &  & or $\alpha_{\rm ov}$ & [cm$^2$\,s$^{-1}$] &  &  \\
  \hline
  1    & Basic        & OPAL & NP12 & 3.20  & 0.71 & 0.020 & 0.693 & 0.000 & --- & 18\,192 & 10.9 \\  
  2    & Composition  & OPAL & NP12 & 3.15  & 0.69 & 0.010 & 0.662 & 0.030 & --- & 11\,402 & 23.3 \\  
  3    & Evolved      & OPAL & NP12 & 3.50  & 0.71 & 0.010 & 0.254 & 0.030 & 2   & 23\,199 & 21.8 \\  
{\bf   4  }  & {\bf Mixing$+$Fine}& {\bf OPAL} & {\bf NP12} & {\bf 3.25}  & {\bf
  0.71} & {\bf 0.014} & {\bf 0.627} & {\bf 0.017} &
 {\bf  1.75} & {\bf 1\,711} & {\bf 1.42} \\  
  5    & Mixture      & OPAL & A09  & 3.18  & 0.71 & 0.021 & 0.640 & 0.018 & 2.00 & 2\,500 & 3.60 \\  
  6    & Mixture      & OPAL & A05$+$Ne& 3.20&0.71 & 0.013 & 0.641 & 0.018 & 2.00 & 5\,905 & 4.53 \\  
  7    & Mixture      & OP   & NP12 & 3.24  & 0.71 & 0.014 & 0.636 & 0.017 & 1.75 & 3\,034 & 5.23 \\ 
  8    & Mixture      & OP   & A09  & 3.22  & 0.71 & 0.015 & 0.638 & 0.018 & 2.00 & 2\,348 & 3.23 \\ 
  9    & Mixture      & OP   & A05$+$Ne& 3.17&0.71 & 0.012 & 0.632 & 0.018 & 1.75 & 3\,249 & 2.90 \\ 
  10   & Step Overshoot& OPAL& NP12 & 3.19  & 0.71 & 0.019 & 0.628 & 0.21  & 1.75 & 3\,792 & 5.54 \\ 
  11   & Metallicity  & OPAL & NP12 & 3.13  & 0.71 & 0.028 & 0.628 & 0.018 & 1.75 & 2\,187 & 2.32 \\ 
  \hline
\end{tabular}
\end{center}
\end{table*}

Models 5 to 9 originate from the Mixture sub-grids.
They provide a test of the influence of different opacities and mixtures on the 
best values for $X_{\rm c}$, $f_{\rm ov}$ and $\log D_{\rm mix}$.
Regardless of their $\chi^2_{\rm red}$ values, all these models converge to very 
similar values for $X_{\rm c}$=0.63 to 0.64, $f_{\rm ov}$=0.017 to 0.018, and 
$\log D_{\rm mix}$=1.75 to 2.00.
However, the mass varies somewhat and the metallicity varies widely over the scanned range.
The fact that $X_{\rm c}$, $f_{\rm ov}$ and $\log D_{\rm mix}$ hardly depend 
on the mass, metallicity, mixture, and opacities manifests that a solid 
seismic constraint is placed on these values for the specific case of KIC\,10526294.
The discretisation of the grid parameters is also sufficiently optimal that the best 
values found are close to one another.

Models 5 and 8 have the A09 mixture and provide the fourth and third best
fits to the frequencies, respectively. 
Although they are built from different opacities, the fit
quality of these two models is indistinguishable in a statistical sense.  Both
are of lower fit quality than Model\,4, given that the difference between their
$\chi^2_{\rm red}$ and the one of Model\,4 is far above 3.84 (which would
correspond to a $p-$value of 0.05 in the case of nested models).
Furthermore, Model\,4 and Model\,7 have identical composition and mixture, but were
calculated using OPAL and OP opacity tables, respectively.  The parameters of
these two models are very close in fit quality, although the latter has the
worse $\chi^2_{\rm red}$.  Models 6 and 9, with the A05$+$Ne mixture, also
provide reasonably good fits to the frequencies.  Therefore, changing the
mixture, at least in this case, influences the adiabatic frequencies more than
swapping between OP and OPAL opacity tables. 

The $\chi^2_{\rm red}$ of Model\,10 is more than twice as 
high as the one of the best model. Hence, we come to the important conclusion
that, for this slowly rotating SPB star the exponentially decaying
prescription for core overshooting gives a better fit to the seismic data 
than a step function.
A more reliable distinction between the two prescriptions for massive pulsators in
general requires a reasonably large ensemble of B-type pulsators to be modelled
with the two prescriptions. A re-analysis of the sample in \cite{aerts-2015}
could be highly instructive in this regard.
Previously, \citet{dziembowski-2008-01} already considered an overshooting
prescription based on two free parameters rather
than a simple step function formulation in their modelling of the $\beta\,$Cep 
stars $\nu\,$Eri and 12\,Lac, but these two pulsators did not allow firm conclusions
on the shape of the overshooting.

Model\,11 is selected from the Metallicity Grid, and is ranked as the second
best asteroseismic model for our target.
Even though it has a different initial mass and metallicity, the overshooting and extra
mixing parameters perfectly agree with those of Model\,4 to Model\,9.
Because Model\,11 has 100\% higher metal abundance than Model\,4, it has a distinct 
stability property that is addressed in Sect.\,\ref{s-excitation}.
Based on our $\chi^2_{\rm red}$ minimisation, Model\,4, and to a lesser extent Models\,11 
and 8, best represent the asteroseismic data of KIC\,10526294.
In the following sections, we use these three models for illustrative purposes.

\subsection{Comparing best models and observations}\label{ss-confront}

\begin{figure*}
\includegraphics[width=\textwidth]{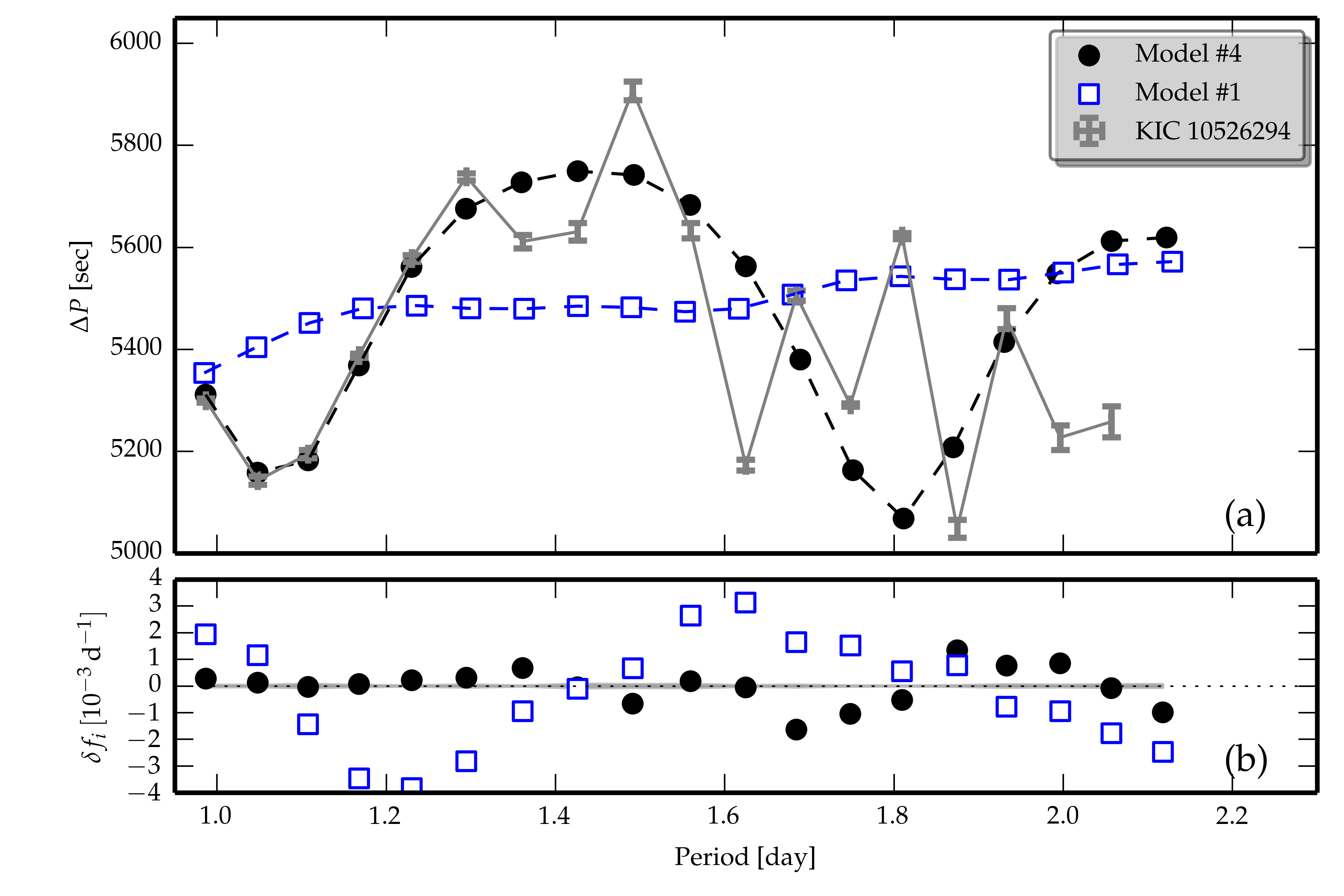}
\caption{(a) Period and period spacings for Model\,4 (filled circles) and Model
  1 (empty squares) from Table\,\ref{t-chisq}.  The observed pattern is shown
  in grey.  (b) The frequency difference between model frequencies and those
  detected in the observations $\delta f_i$ from Eq.\,(\ref{e-delta-f}).
  Compared to Fig.\,\ref{f-obs-dP}b, the ordinate is enlarged 40 times; the grey
  band around zero is the 1$\sigma$ frequency uncertainty range $\sigma_i$.  A
  similar plot for Model\,4, Model\,8 and Model\,11 is shown in 
  Figs.\,\ref{f-dP-4-vs-11} and \ref{f-dP-8-vs-11} in the Appendix.
\label{f-best-chisq}}
\end{figure*}

Fig.\,\ref{f-best-chisq}a compares the observed period spacing (grey symbols) 
versus mode periods with those of Model\,4 (filled circles) and Model\,1 of P14 
(empty squares), respectively.
The resulting period spacing pattern of Model\,4 clearly follows the observed pattern
better than the pattern from Model\,1 does.
To appreciate how well 
the frequencies from Model\,4 approach the observed ones, we show
in Fig.\,\ref{f-best-chisq}b the frequency deviation $\delta f_i$ 
between the two through
\begin{equation}\label{e-delta-f}
\delta f_i = f_i^{\rm (th)} - f_i^{\rm (obs)}, \quad i=1, \,\cdots, \,19.
\end{equation}
In Fig.\,\ref{f-best-chisq}b, the ordinate range is 40 times larger than in
Fig.\,\ref{f-obs-dP}b, and the 1$\sigma$ frequency uncertainty band appears as a
line here.  The frequencies of Model\,1 (empty squares) show stronger
deviations from zero and the frequencies of Model\,4 provide far better fits
to the observed period series, especially in the lower period regime.
To appreciate the quality of the model fit, we computed the average 
$\delta f_i$ for Model\,4 and obtained 0.00051\,d$^{-1}$.
While this is still a factor $\sim30$ larger than the observational uncertainty
on the frequencies, it is by far the best seismic model constructed to
represent the pulsations of a late B-type star up to now.

Additional comparisons of the period spacing fits between Models\,4 and 11 and
between Models\,8 and 11, are presented in Fig.\,\ref{f-dP-4-vs-11} and 
Fig.\,\ref{f-dP-8-vs-11} in the Appendix.
We list the global parameters of the three best seismic models of KIC\,10526294
in Table\,\ref{t-Model-4}.  
Their structure variables (in GYRE-compatible format) are available online; see Appendix\,\ref{s-deliv}.
Even for optically bright B-type pulsators, the seismically derived
  effective temperature and surface gravity do not necessarily agree with their
  spectroscopic counterparts because the latter suffer from systematic
  uncertainties \citep[e.g.][]{briquet-2007-01,briquet-2011-01}.  In the case of
  KIC\,10526294, which is a faint star, the seismic values of the best models
  all agree with the spectroscopic values within 3$\sigma$, which we regard as
  good compatibility between seismology and spectroscopy.

Fig.\,\ref{f-corr-planes} shows the correlation diagrams between the five
parameters of the Mixing Grid and the Fine Grid centred around the
parameters of Model\,4, that is, all panels present a subsurface in the
$\chi^2_{\rm red}$ space where the parameters that are not shown on the axes
are fixed to the values found for Model\,4.  This is similar to Fig.\,11 in
P14, where semi-linear trends between all parameters occur.  Here, we
achieved more stringent constraints on some of the parameters, especially
$X_c$, $f_{\rm ov}$, and $\log D_{\rm mix}$ (see also
Fig.\,\ref{f-corr-scatter}).  
On the other hand, $M$ and $Z$ are less well constrained, although we achieved a 
relative precision below 4\% for the mass.
A strong linear
correlation occurs between them, cf.\ panel (e), which is similar to Fig.\,2 in
\citet{ausseloos-2004-01}.  For a full overview of the $\chi^2_{\rm red}$
distribution for the entire Mixing Grid, Fine Grid and Metallicity Grid, we refer to
Fig.\,\ref{f-corr-scatter} in the Appendix.
The existence of correlations between the parameters hinders the derivation
  of appropriate uncertainties on the derived parameters of the best model(s),
  other than the bare minimum uncertainty given by the parameter stepsizes. A
 pragmatic estimate of the overall systematic and statistical uncertainty is
  provided by the ranges of the parameters of the best three models listed in 
Table\,\ref{t-Model-4}.

\begin{table}
\begin{center}
	\caption{Global parameters of Model\,4, Model\,8, and Model\,11.
		The free grid parameters and the derived model parameters
        are separated by the horizontal line.
		$d_{\rm ov}$ and $M_{\rm ov}$ are the width and the mass contained 
		in the overshooting zone, respectively.}
	\label{t-Model-4}
	\begin{tabular}{llll}
		\hline
		Parameter                & Model\,4      & Model\,8 & Model\,11 \\
		\hline
		Mass [M$_\odot$]         & 3.25            & 3.22  & 3.13  \\
		$f_{\rm ov}$             & 0.017           & 0.018 & 0.018 \\
		$X_{\rm c}$              & 0.627           & 0.638 & 0.628 \\
		$X_{\rm ini}$            & 0.710           & 0.710 & 0.710 \\
		$Y_{\rm ini}$            & 0.276           & 0.275 & 0.262 \\
		$Z_{\rm ini}$            & 0.014           & 0.015 & 0.028 \\
		$\log D_{\rm mix}$ [cm$^2$\,s$^{-1}$] & 1.75 & 2.00& 1.75 \\
		\hline
		Radius [R$_\odot$]       & 2.215           & 2.195 & 2.382 \\
		Luminosity [L$_\odot$]   & 128             & 111   &  78   \\
		T$_{\rm eff}$ [K]        & 13\,000         & 12\,650 & 11\,100  \\
$\log\,g$ (cgs) & 4.259 & 4.263 & 4.204 \\
		Age [Myr]                & 63.0            & 61.9  & 91.7  \\
		$M_{\rm cc}$ [M$_\odot$] & 0.672           & 0.651 & 0.591 \\ 
		$R_{\rm cc}$ [R$_\odot$] & 0.329           & 0.326 & 0.318 \\  
		$M_{\rm ov}$ [M$_\odot$] & 0.190           & 0.195 & 0.178 \\ 
		$d_{\rm ov}$ [R$_\odot$] & 0.037           & 0.038 & 0.037 \\ 
		\hline
	\end{tabular}
\end{center}
\end{table}

Based on Models\,4, 8 and 11, we deduce that KIC\,10526294 
is a young star ($X_{\rm c}\approx0.63$ to 0.64) of late-B spectral type.
From the parameters of Models\,4 to 9 and Model\,11 in Table\,\ref{t-chisq},
the core overshoot parameter is tightly constrained in the range 
$f_{\rm ov}=0.017$ to 0.018. 
The overshooting zone has enough spatial and mass extension to allow partial 
trapping of some of the pulsation modes.
This offers a unique opportunity for an in-depth study of the properties of the
overshooting layer.  
Additionally, an extra diffusive mixing in the radiative part of the star of an 
amplitude $\log D_{\rm mix}=1.75$ to 2.00 cm$^2$\,s$^{-1}$ is needed to explain 
the observed frequencies.
The parameters 
$\log D_{\rm mix}$, $f_{\rm ov}$, and $X_{\rm c}$ only show a minor dependence on 
the adopted mixture and choice of the opacity tables.

\begin{figure*}
	\includegraphics[width=\textwidth]{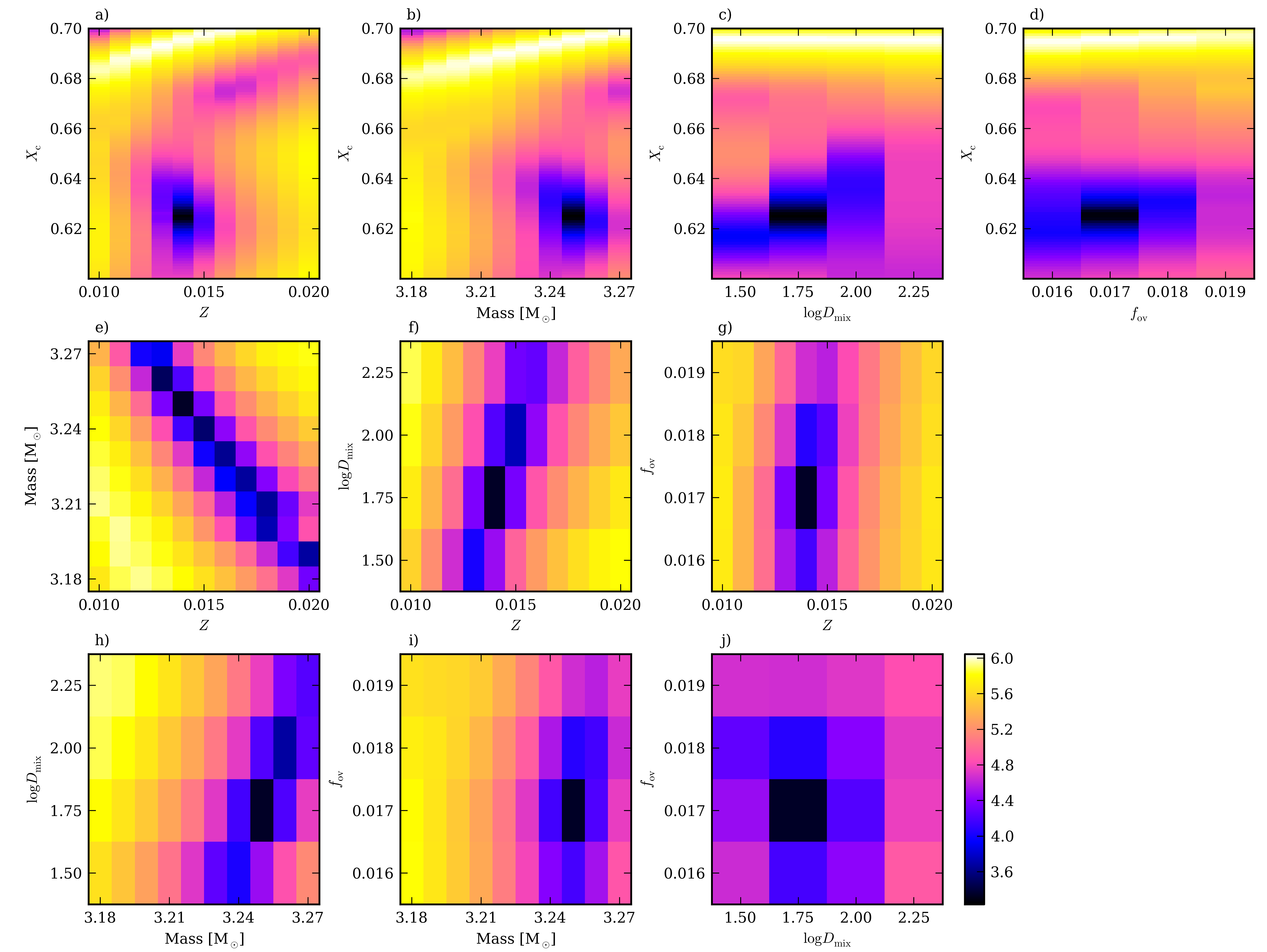}
	\caption{Correlation diagram of the parameters of the Fine Grid around
          Model\,4.  Results are based on the Mixing Grid plus the Fine
          Grid.  The colour coding is based on $\log\chi^2_{\rm red}$
          {\color{blue}}.  To see the improvement we achieved in confining the grid
          parameter space, compare with Fig.\,11 in P14. In each panel, the
          model parameters that are not on the axis are fixed to those of
          Model\,4.  Here, $X_{\rm c}$, $f_{\rm ov}$ and $\log D_{\rm mix}$
          are well constrained.}
	\label{f-corr-planes}
\end{figure*}

\begin{figure}
	\includegraphics[width=\columnwidth]{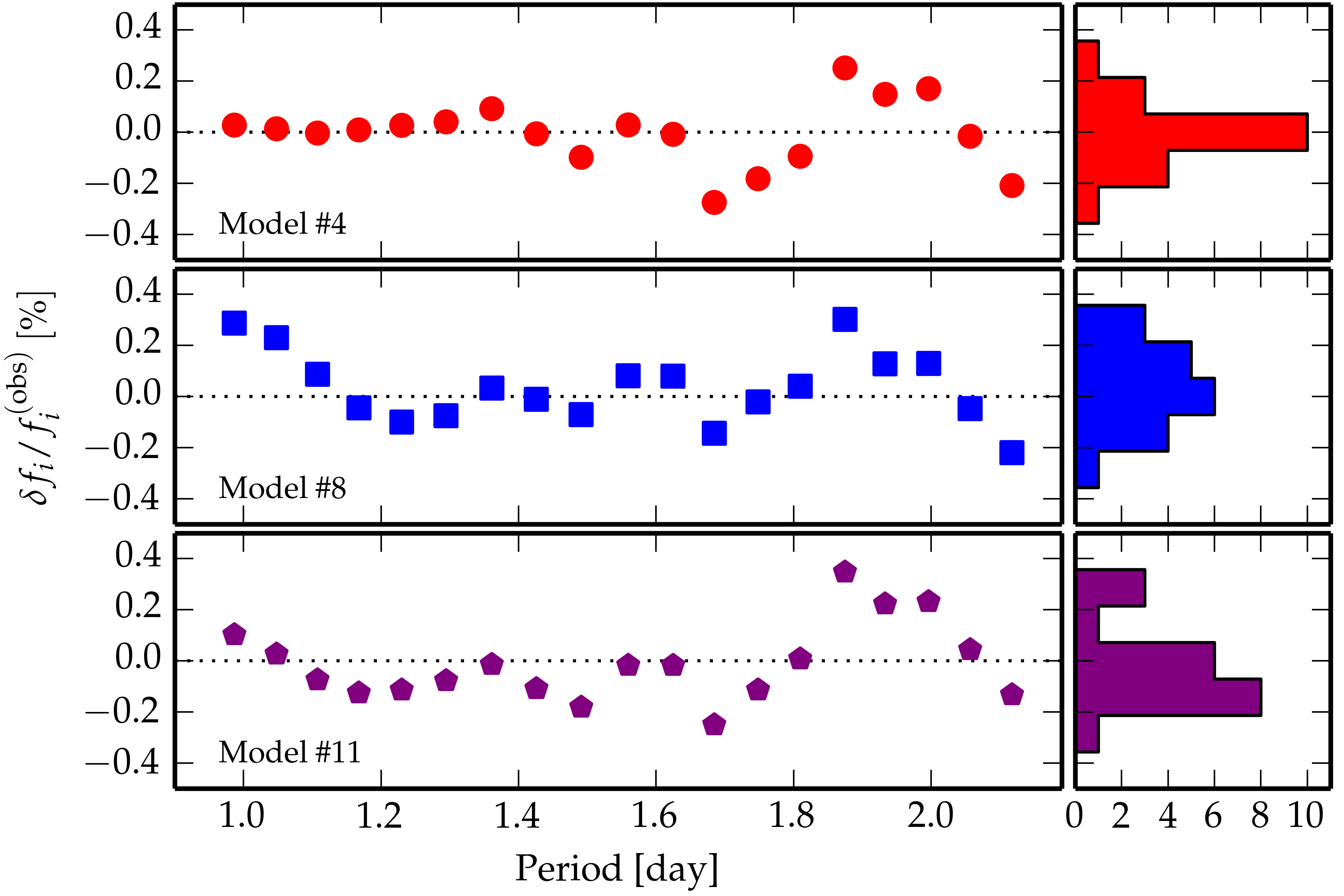}
	\caption{Left panel: Relative deviation between observed and modelled frequencies 
	   $(f^{\rm (th)}_i - f^{\rm (obs)})/f^{\rm (obs)}$ for the 19 dipole
	   g-modes in KIC\,10526294.
	   The deviation for all models is below 0.4\%.}
	   Right panel: Histogram distribution of the relative deviations around zero.
	\label{f-rel-dev}
\end{figure}

One may question the success of our forward modelling for KIC\,10526294,
specifically given the high $\chi^2_{\rm red}$ scores of the best model(s) in
Table\,\ref{t-chisq}.  We argue that despite such high values, the
relative frequency deviations between the observed frequencies and
those of the best seismic model remain below 0.4\%.  The left panels in
Fig.\,\ref{f-rel-dev} show $\delta f_i/f^{\rm (obs)}_i$ versus mode period
$1/f^{\rm (obs)}_i$ for Model\,4 (top) Model\,8 (middle) and Model\,11
(bottom), respectively.  The strongest deviations occur for modes with periods
exceeding $\sim1.68$ days, that is, $f_1$ to $f_8$ in Table\,\ref{t-freq}.  For
Model\,4, the first eleven mode frequencies are very similar to 
the observed ones; the
deviations gradually grow by increasing mode order.  Current seismic
models of solar-like stars observed with {\it Kepler\/} give rise to relative deviations
of about 10$^{-4}$ between observed and modelled frequencies near the frequency
of maximum power \citep{metcalfe-2014-01}, but this is only achieved after applying an
arbitrary surface correction \citep{kjeldsen-2008-01}. 
Our results are only an order of magnitude worse, but we are
treating a star very different from the Sun; 
this is a major achievement. 
For completeness, we point out that we did not find any correlation between the mode
amplitudes and the relative frequency deviations shown in Fig.\,\ref{f-rel-dev}.

The right panels in Fig.\,\ref{f-rel-dev} show histograms of the relative
frequency deviations between the measurements and theoretical predictions around
zero.  Despite the low-number statistics, the deviations for Model\,4 are
normally distributed, while those of the other two models do not show a
  clear Gaussian shape.  
For the solar-type pulsators, a frequency-dependent correction is applied to the 
theoretical frequencies to account for
near-surface effects \citep[e.g.][]{kjeldsen-2008-01}. For B-type
pulsators, such a correction has so far been assumed to be unnecessary because these
stars have a radiative envelope.  If such surface a correction were needed for
B-type stars, then we would have observed a monotonically deviating behaviour
between the model frequencies and the observed ones, which is clearly not the
case.  Therefore, we argue that even if a surface correction term would be needed
for KIC\,10526294, it would be frequency-independent.

\begin{figure}
	\includegraphics[width=\columnwidth]{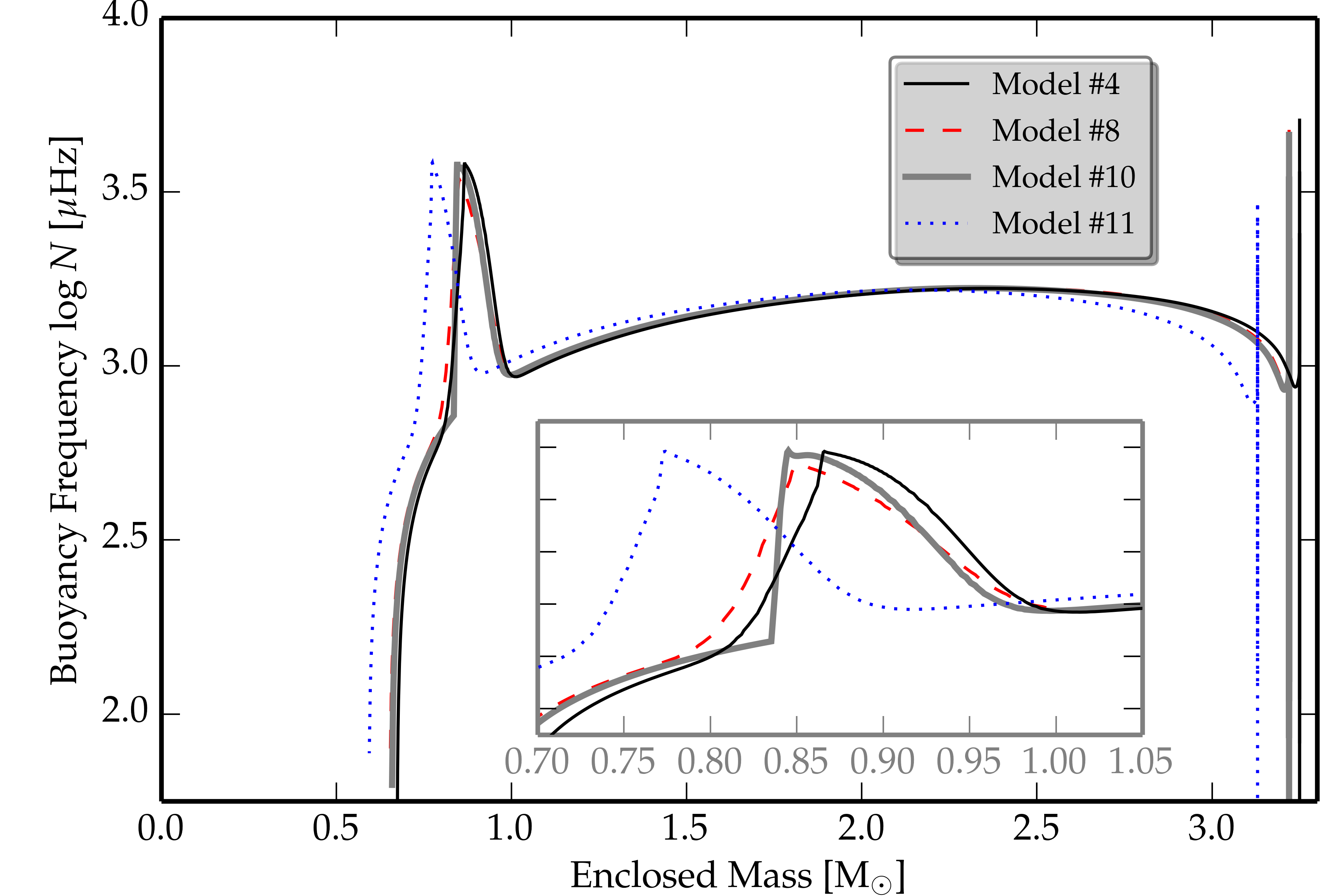}
	\caption{Brunt-V{\"a}is{\"a}l{\"a} frequency for Model\,4 (black solid),
	Model\,8 (red dashed), Model\,10 (grey solid) and Model\,11 (blue dotted) 
	from Table\,\ref{t-chisq} and Table\,\ref{t-Model-4}.
	The inset is a zoom-in around the peak in the profile induced by the gradient of
	the mean molecular weight outside the convective core.
	These four models have different masses of the convective core 
    M$_{\rm cc}$.}
	\label{f-brunt}
\end{figure}

Fig.\,\ref{f-brunt} compares the profile of the Brunt-V{\"a}is{\"a}l{\"a} frequency 
$N(r)$, as used in Eq.\,(\ref{e-dP}) for four selected models.
The inset is a zoom-in around the sharp increase in the profile, induced by the 
gradient of the mean molecular weight $\nabla_\mu$ outside the receding core.
The profiles for Model\,4 (black solid), Model\,8 (red dashed) and Model\,11
(blue dotted) are almost identical, because of their very similar overshooting parameter
$f_{\rm ov}=0.017$ to 0.018, except for a slight shift in the position of the peak
that is due to their different initial masses.
The profile for Model\,10 (grey solid) shows a steeper rise around 0.85\,M$_\odot$
because of the difference between the exponentially decaying and step function 
prescriptions for overshooting.
The seismic frequency fitting indicates that the smoother 
shape of $N(r)$ connected
with the exponentially decaying prescription is to be preferred
over the steep step-function for KIC\,10526294.

\section{\label{s-excitation}Mode excitation}

In their Fig.\,14, P14 already showed a mismatch between the predicted and
observed excited modes using their best model (i.e. Model\,1 here).  Only
eleven modes out of nineteen in Model\,1 are excited.  Aside from the forward
modelling, we examined if our improved modelling remedies that shortcoming.  For
this, we used the non-adiabatic framework of GYRE (version 3.2.2).

It is well known that current predictions of mode excitation through the heat
mechanism are not yet sufficiently appropriate to explain all the
detected and identified modes in B-type stars \citep{dziembowski-2008-01}.  For
this reason, we examined the mode stability properties of our best models
a posteriori.

\begin{figure*}
\begin{minipage}{0.48\textwidth}
\includegraphics[width=\columnwidth]{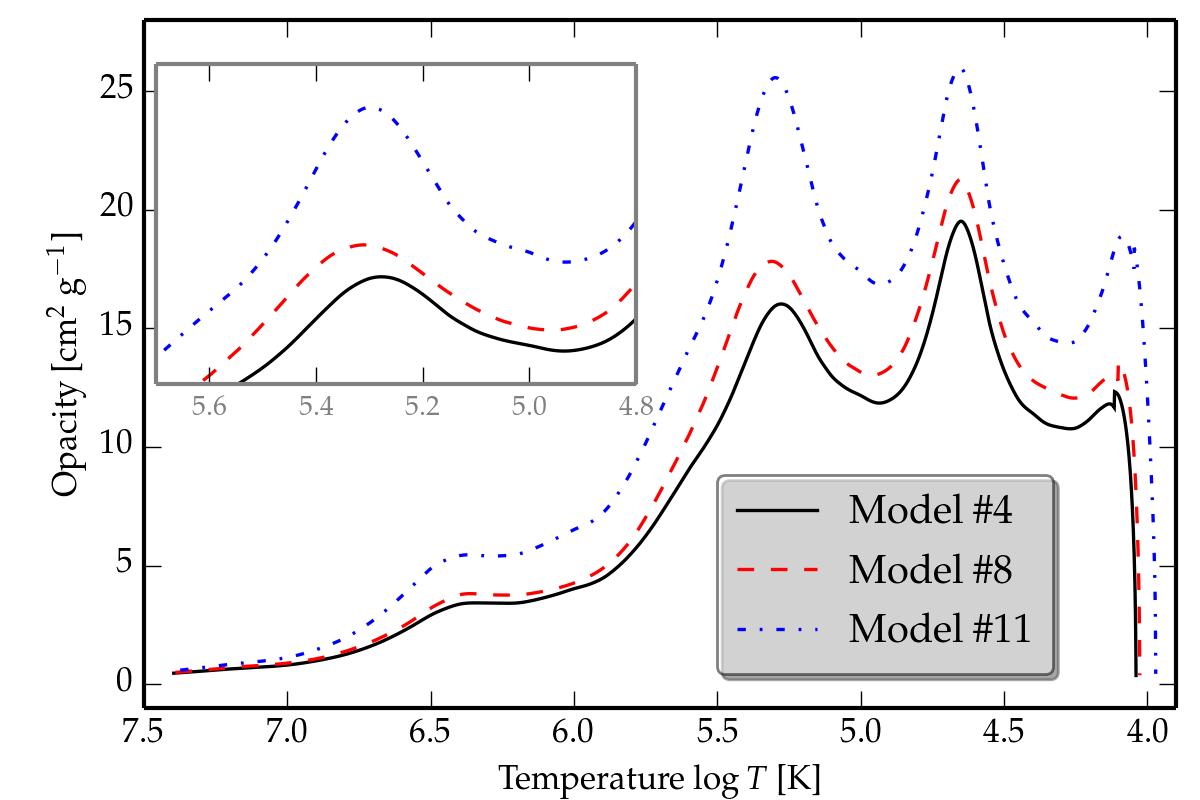}
\caption{Rosseland mean opacity profile in Model\,4 ($Z_{\rm ini}=0.014$, 
OPAL, NP12, black solid line), Model\,8 ($Z_{\rm ini}=0.015$, OP, A09, red dashed-dotted line)
and Model\,11 ($Z_{\rm ini}=0.028$, OPAL, NP12, blue dashed line)  versus temperature.
The inset is a zoom around the iron-bump of opacity.
The iron opacity peak for the OP model occurs at slightly hotter interior.}
\label{f-kappa}
\end{minipage}
\hspace{0.04\textwidth}
\begin{minipage}{0.48\textwidth}
\includegraphics[width=\columnwidth]{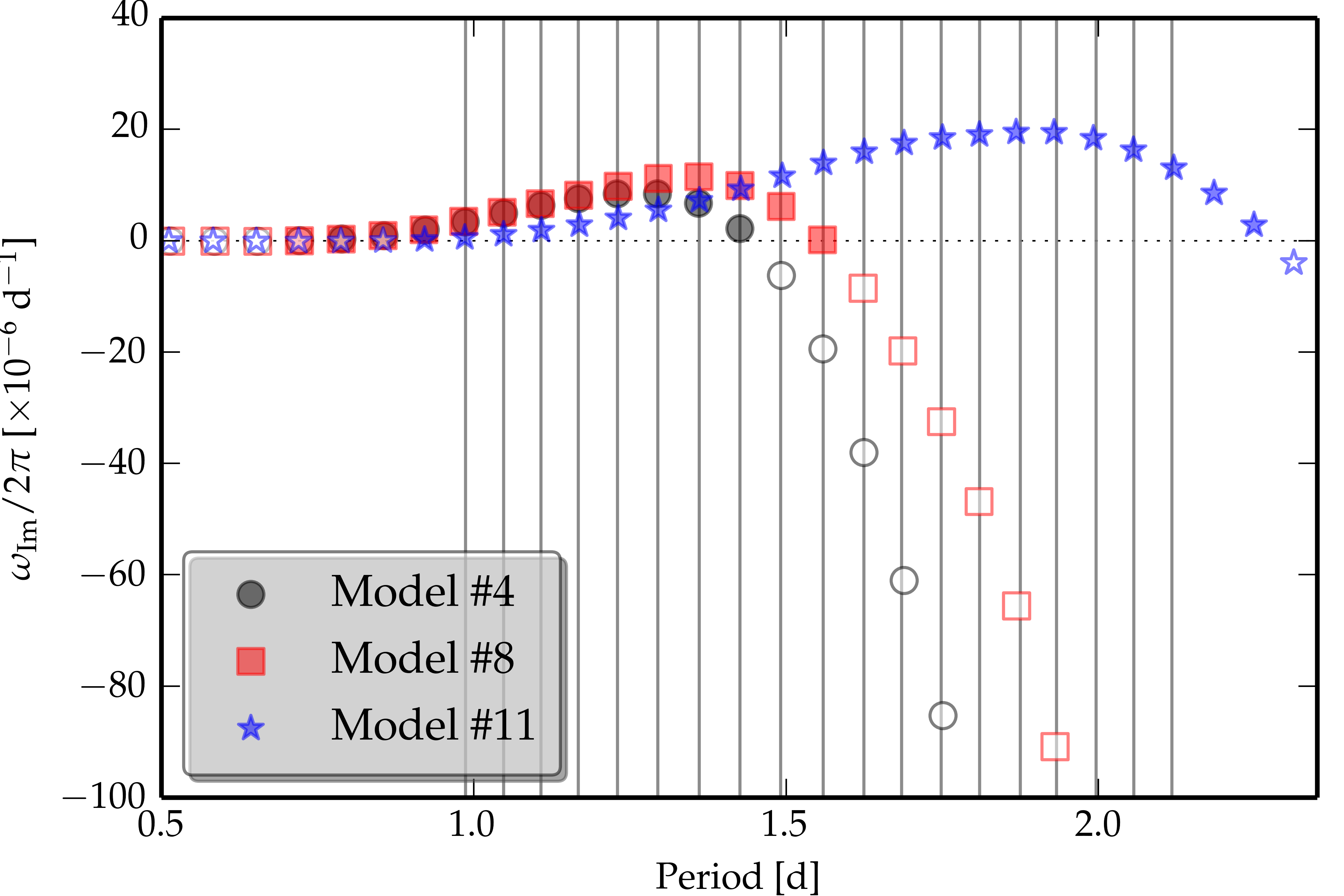}
\caption{Non-adiabatic instability study of Model\,4 (black circles), 
Model\,8 (red squares), and Model\,11 (blue stars) from Table\,\ref{t-chisq}.
The imaginary part of the eigenfrequency $\omega_{\rm Im}$ is plotted against the mode period
(vertical lines).
Filled (empty) symbols designate excited (damped) modes.}
\label{f-excitation}
\end{minipage}
\end{figure*}

\citet{miglio-2007-01} already suggested that adopting OP opacity tables instead
of OPAL might remedy the mode excitation problem. As a result of the significant role
of the opacity profile in mode excitation, we first considered its profile in
three of the best selected models.  Fig.\,\ref{f-kappa} compares the Rosseland
mean opacity profile $\kappa$ versus temperature $\log T$ in Model\,4
(black solid line, using OPAL), Model\,8 (red dashed line, using OP) and
Model\,11 (blue dashed-dotted line, using OPAL).  The inset is a zoom in of the Fe
opacity bump.  Obviously, Model\,11, with its twice as high initial
metallicity $Z_{\rm ini}$ is more opaque than the other two.  Moreover,
Model\,8 is more opaque than Model\,4 despite their nearly identical $Z_{\rm
  ini}$, because the former model uses the OP opacities, in line with
\citet{miglio-2007-01}.

In Fig.\,\ref{f-excitation}, we compare the growth rate (i.e. the imaginary part 
of the complex eigenfrequency $\omega_{\rm Im}$) as a function of mode period for 
Model\,4 (black circles), Model\,8 (red squares) and Model\,11 (blue stars), 
respectively. 
The vertical lines are the observed periods. 
The excited (damped) modes are shown with filled (empty) symbols.
Although Model\,4 reproduces the real part of eigenfrequencies 
better than Model\,11, the latter predicts more excited modes than observed.
This is not a surprise given the high $Z_{\rm ini}$ of Model\,11, which 
implies a $\sim70\%$ higher iron opacity (see Fig.\,\ref{f-kappa}).
We point out that an ad-hoc local increase in the height of the iron opacity 
bump by up to $\sim50\%$ \citep[cf. ][]{pamyatnykh-2004-01}
would allow Model\,4 (as well as other models from the Mixing Grid and 
Fine Grid) to have equally good mode excitation properties as Model\,11.
Model\,8 has two additional excited modes compared to those of Model\,4, despite both having
similar $Z_{\rm ini}$.
This is as expected because OP opacities predict more excited modes \citep{miglio-2007-01}.
Additionally, all three models presented here predict two to four short-period modes to be 
excited but these are not among the detected modes listed in P14.

The distinction between Model\,4, Model\,8 and Model\,11 is not
  clear-cut; the former offers the best adiabatic frequency match with the
  observations, while the latter explains the mode excitation better at the
  cost of a somewhat poorer frequency fitting.  Additionally, the metallicity
  value $Z_{\rm ini}$ of Model\,4 and of Model\,8 perfectly agrees with the
  Galactic standard composition of \citet{nieva-2012-01} for nearby B stars,
  while Model\,11 is twice as metal-rich as the standard.  Given the current
  lack of a higher signal-to-noise spectrum in addition to the one presented in
  P14 upon which we relied, we cannot evaluate this metallicity difference
  between the best models from spectroscopic abundance analysis. Fortunately,
  this degeneracy does not affect the seismic estimates of the core overshooting
and diffusive mixing deduced for the star.

\section{Discussion and conclusions}
\label{s-discussion}

Forward seismic modelling based on the 19 detected and identified dipole gravity
modes of the B8.3V star KIC\,10526294 led to the derivation of its mass, radius,
and age 
within the ranges $M\in [3.13,3.25]\,$M$_\odot$, $R\in
  [2.19,2.38]$\,R$_\odot$, and age $[62,92]$\,Myr, respectively.  These ranges
  cannot be refined to tighter constraints as long as we do not have a
  constrained metallicity from spectroscopy. In practice, the range of $Z\in
  [0.014,0.028]$ leads to good seismic models in this mass range, and the
  $(M,Z)$ correlation implies a quite large uncertainty on the seismic age.
  However, the seismic data do require the inclusion of a small but significant
  amount of extra diffusive mixing in the radiative envelope of the star, with a
  value of $\log D_{\rm mix}$ in the range $[1.75,2.00]$\,cm$^2$\,s$^{-1}$, in
  addition to an exponentially decaying core overshooting with parameter 
  $f_{\rm ov}\in [0.017,0.018]$. The tight constraints on $\log D_{\rm mix}$ and
  $f_{\rm ov}$ do not depend on the choice of the mass, metallicity, mixture and 
  opacity table within the allowed seismic ranges. We also conclude that models 
  with a step function core overshooting description are of inferior quality 
  to explain the seismic data of this star.

Our work represents the first application of seismic modelling of an SPB
  from which the need of global extra mixing is proven and quantified with high
  precision, based on the observed frequencies of unambiguously identified modes.
  Addressing the origin of the physical mechanisms that generate the desired
  value of $D_{\rm mix}$, and their possible interaction(s), is yet to be done,
  keeping in mind that KIC\,10526294 is an ultra-slow rotator.

Forward seismic modelling is currently the best starting point to describe the
global physical parameters, the interior structure characteristics, and the
detailed pulsational properties of real stars.  To carry out an iterative
asteroseismic inversion for the rotation profile or for the entire structure of
the star, one has to start from a calibrated model that has the ability to fit
all the detected and identified oscillation modes reasonably close to their
measured frequencies.  In this study, we have achieved this for the slowly
rotating \textit{Kepler\/} target KIC\,10526294.

Despite the still relatively high $\chi^2_{\rm}$ values for the best model(s),
Fig.\,\ref{f-rel-dev} shows that the relative deviation between the modelled 
and observed frequencies is below 0.4\%. 
Incorporating non-adiabatic corrections to the adiabatic frequencies, 
and moving beyond the linearised oscillation paradigm 
\citep[e.g.][]{vanhoolst-1994-01} may further reduce the
deviations between the observed and model frequencies.  In addition, there is
still room for improving the mode stability predictions for KIC\,10526294
because 
(a) depending on the adopted metallicity, several of the observed longest-period 
dipole modes are predicted to be stable, and 
(b) few excited dipole modes are not observed.

Thanks to the unprecedented precision of the
observations assembled by the nominal \textit{Kepler\/} mission, we have now
reached the stage at which we can employ asteroseismic methods to calibrate the internal 
structure of massive stars with well-developed convective cores, and 
quantify the level of seismic modelling.
This allows us to evaluate different choices of the input physics of B-star 
models coupled with an appropriate statistical framework (such as the 
$\chi^2_{\rm red}$) to properly account for the different of the models.
Our modelling of KIC\,10526294 proves that we can adequately explain the high-precision 
data collected from space within the framework of spherically
symmetric stellar evolution models, and a linearised formalism of stellar
oscillations below percent level.

\begin{acknowledgements}
  We are grateful to the valuable comments from the anonymous referee.
  EM is beneficiary of a postdoctoral grant from the Belgian Federal Science
  Policy Office, co-funded by the Marie Curie Actions FP7-PEOPLE-COFUND-2008
  n$^\circ$246540 MOBEL GRANT from the European Commission.  Part of the
  research included in this manuscript was based on funding from the Research
  Council of KU\,Leuven under grant GOA/2013/012 and from the National Science
  Foundation of the USA under grant No.\,NSF PHY11-25915.  The computational
  resources and services used in this work were provided by the VSC (Flemish
  Supercomputer Center), funded by the Hercules Foundation and the Flemish
  Government – department EWI.  EM and CA thank Bill Paxton, Richard Townsend,
  and Francis Timmes for their valuable support with the MESA and GYRE codes, 
  Steven Kawaler for fruitful discussions about period spacings, and
  Andrea Miglio for various discussions on extra diffusive mixing over the past
  years.  CA and PIP acknowledge the staff of the Kavli Institute of Theoretical
  Physics at the University of California, Santa Barbara, for the kind
  hospitality during the 2015 research programme ``Galactic Archaeology and
  Precision Stellar Astrophysics''.  EM is grateful to the organisers of IAU\,307
  symposium in Geneva where part of this work was presented.
\end{acknowledgements}

\bibliographystyle{aa}
\bibliography{my-bib}


\begin{appendix}

\section{Period spacing for lternative models}
\label{s-other-models}

\begin{figure}
	\includegraphics[width=\columnwidth]{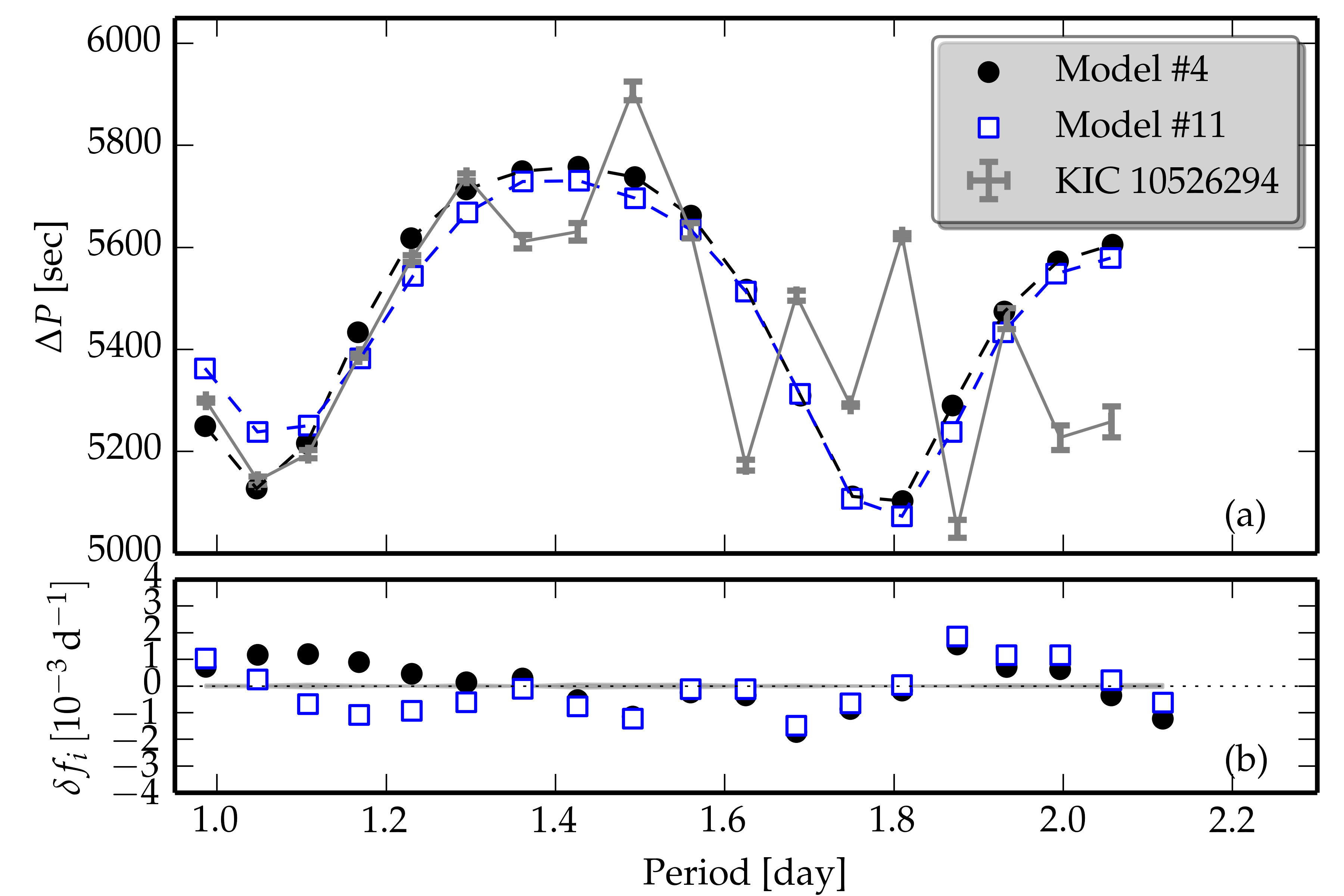}
	\caption{Top panel: Period spacing for Model\,4 (filled circles) and
          Model\,11 (empty squares).  See Table\,\ref{t-chisq} for their
          parameters.  Bottom panel: Absolute frequency deviations of the two models
          with respect to observations.  See also Fig.\,\ref{f-best-chisq} for a
          comparison.}
	\label{f-dP-4-vs-11}
\end{figure}

\begin{figure}
	\includegraphics[width=\columnwidth]{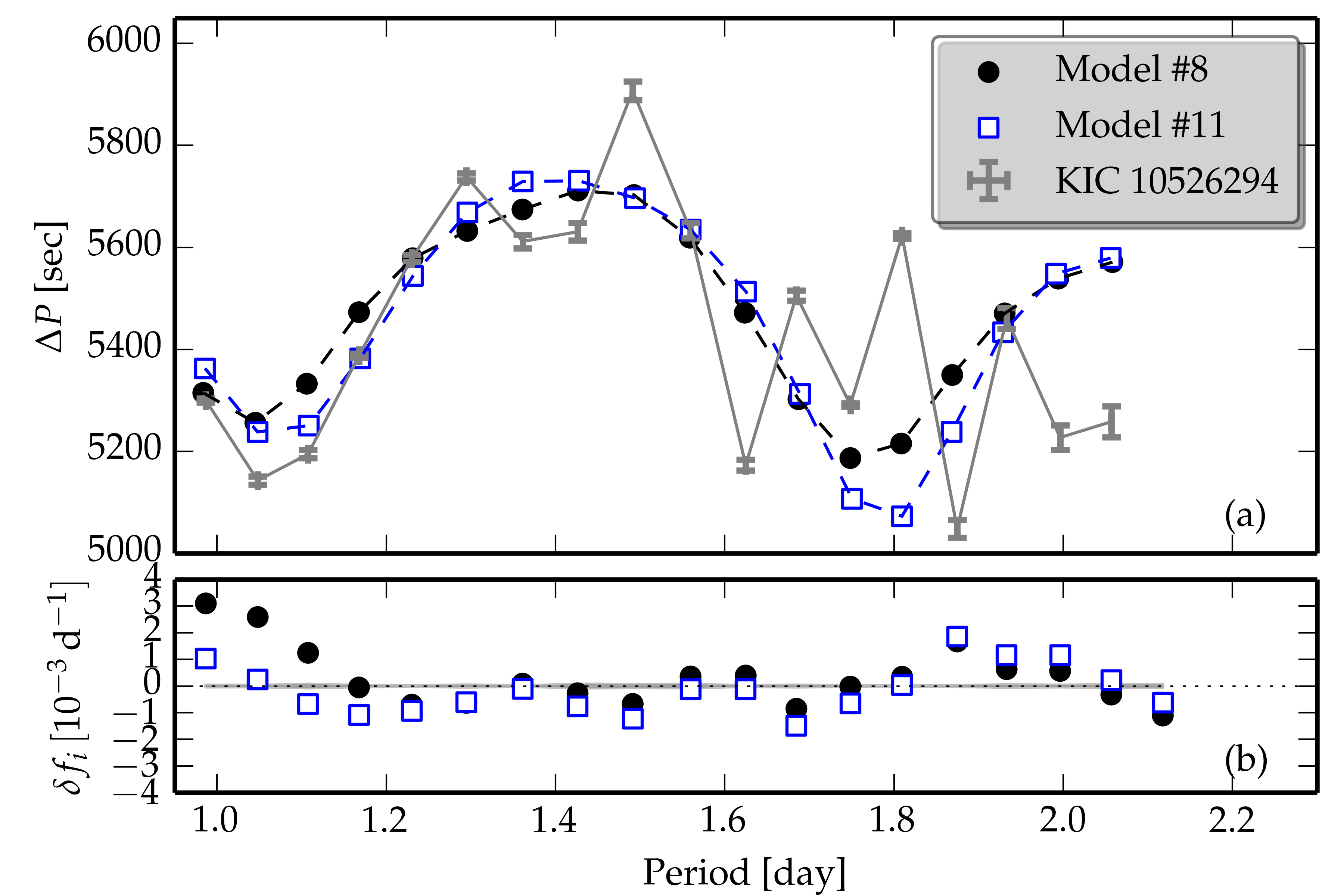}
	\caption{Similar to Fig.\,\ref{f-best-chisq} but for Model\,8 (filled circles) 
	      and Model\,11 (empty squares).}
	\label{f-dP-8-vs-11}
\end{figure}

The comparison between the period spacing of the first and second best models is
presented in Fig.\,\ref{f-dP-4-vs-11}.
A similar plot for the second and third best models is shown in Fig.\,\ref{f-dP-8-vs-11}.
Both Model\,8 and Model\,11 provide poorer fits to the period spacing on the short-period 
range of the series, than Model\,4 (Fig.\,\ref{f-best-chisq}).


\section{Full overview of the Mixing, Fine and Metallicity Grids}
\label{s-corr-scatter}

\begin{figure*}
	\includegraphics[width=\textwidth]{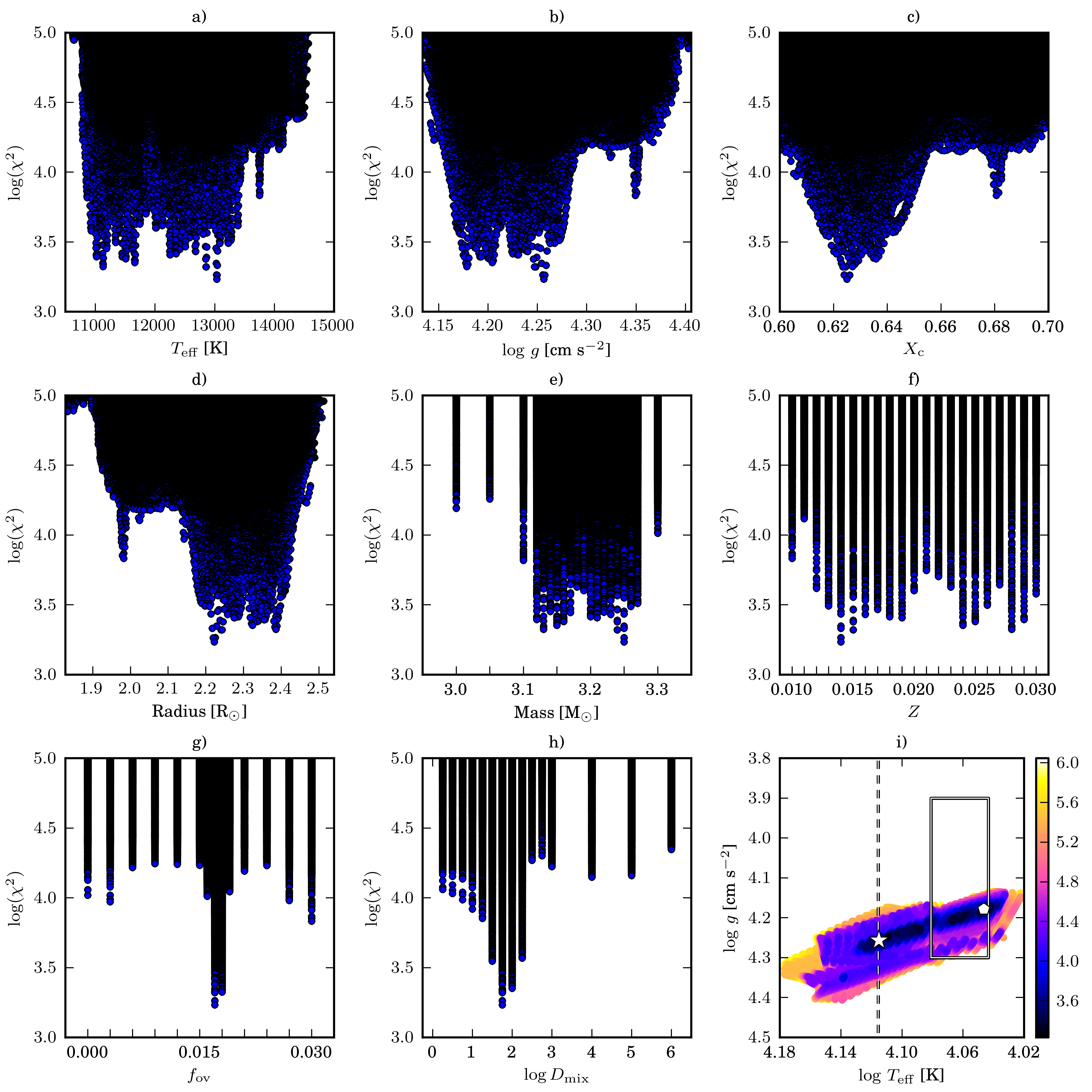}
	\caption{Behaviour of $\log \chi^2_{\rm red}$ as a function of (a)
          effective temperature $T_{\rm eff}$, (b) surface gravity $\log g$, (c)
          age $X_{\rm c}$ (d) radius, (e) initial mass, (f) metallicity $Z$, (g)
          exponential overshooting parameter $f_{\rm ov}$, and (h) extra
          diffusive mixing $\log D_{\rm mix}$.  Panel (i) shows the position of
          the two best models (star: Model 4, pentagon: Model 11) in the Kiel
          diagram, where the box indicates the 1$\sigma$ and the dotted line 
          the 3$\sigma$ boundaries deduced from spectroscopy.  The colour
          coding is based on $\log \chi^2_{\rm red}$.}
	\label{f-corr-scatter}
\end{figure*}

The Mixing, Fine and Metallicity grids introduced in Table\,\ref{t-grid} have
identical physical ingredients, and it is safe to merge their $\chi^2_{\rm red}$
goodness-of-fit into a single snapshot.  Fig.\,\ref{f-corr-scatter} shows
$\log\chi^2_{\rm red}$ as a function of the dimensions of the three grids in
addition to the model radius and surface gravity.  
Note the finger-like
structures in panels (c), (g), and (h) where the age, overshooting and mixing
parameters are constrained, respectively.  In panel (i), the position of all
models is shown on the Kiel diagram ($\log T_{\rm eff}$ vs. $\log g$), and that
of the best model is marked with a (white) star.  Our Model\,4 is found
marginally inside the 2$\sigma$ uncertainty box from spectroscopy.

\section{Deliverables, inlists and opacity tables}
\label{s-deliv}

We adhere to the MESA code of conduct as stated in \citet{paxton-2011-01}, and make our setting files
and physical ingredients publicly available for download.
This also ensures reproducibility of our results.
The following items are available online:
\begin{itemize}
	\item The MESA v.5548 input inlists,
	\item The GYRE v.3.0 inlist,
	\item The OP and OPAL opacity tables adapted to the A05$+$Ne, A09, and NP12 mixtures.       
	They have to be used with a standard MESA composition option \texttt{initial\_zfracs = 5,
	6} and \texttt{8}, respectively.
	All tables are MESA compatible.
	When querying the OP and OPAL servers, we made a choice to redistribute the abundance residuals
	on all metals, based on their relative mass fraction instead of depositing them on the heaviest
	metals, which are Fe and Ni.
	This is a choice and may slightly influence the adiabatic and non-adiabatic results.
	\item The internal structure of Model\,4, Model\,5, Model\,8, and Model\,11 
	in a GYRE-compatible format. 
\end{itemize}
Static links to download each of these products are available from
\href{https://fys.kuleuven.be/ster/Projects/ASAMBA}{https://fys.kuleuven.be/ster/Projects/ASAMBA}.

\end{appendix}

\end{document}